\documentclass[aps,twocolumn,reprint,prd,preprintnumbers,nofootinbib]{revtex4-1}
\pdfoutput=1
\usepackage{amsmath,amssymb,amsthm,tikz,tikz-cd,multirow,bm,graphicx,float,array,rotfloat,pgf,soul,bm,xcolor,comment,svg,stackengine,ragged2e}

\usepackage[all]{xy}
\usepackage[shortlabels]{enumitem}
\usepackage[normalem]{ulem}
\usepackage[hidelinks]{hyperref}
\usepackage[vcentermath]{youngtab}
\usepackage[boxsize=.8em]{ytableau}
\usepackage{todonotes}
\usepackage{dynkin-diagrams}
\usepackage{tikz}
\usepackage{float}
\usepackage{placeins}
\allowdisplaybreaks
\tikzset{Rightarrow/.style={double equal sign distance,>={Implies},->},
triple/.style={-,preaction={draw,Rightarrow}},
quadruple/.style={preaction={draw,shorten >=0pt},shorten >=1pt,-,double,double
distance=0.2pt}}
\usetikzlibrary{positioning}
\usepackage{graphicx}
\usepackage{caption}
\usepackage{subcaption}
\usepackage{comment}
\usepackage{subfiles}
\usepackage{todonotes}
\usepackage{booktabs}
\usepackage{lipsum}
\usetikzlibrary{hobby,intersections} 
\usetikzlibrary{shapes.geometric}
\usetikzlibrary{shapes.misc}
\usetikzlibrary{decorations.pathreplacing}
\def\ns#1#2{
	\node[circle, draw, fill=white] (#2) at (#1){};
	\node[cross out, draw] at (#1){};
}

\tikzset{flavour/.style={draw=none,minimum size=0.3mm,fill=white, regular polygon,regular polygon sides=4,draw}}
\tikzset{flavourr/.style={draw=none,minimum size=0.3mm,fill=red, regular polygon,regular polygon sides=4,draw}}
\tikzset{flavourb/.style={draw=none,minimum size=0.3mm,fill=blue, regular polygon,regular polygon sides=4,draw}}
\tikzset{flavouro/.style={draw=none,minimum size=0.3mm,fill=orange, regular polygon,regular polygon sides=4,draw}}
\tikzset{gaugeBig/.style={inner sep=1mm,draw=none,fill=white,minimum size=2mm,circle, draw}}
\tikzset{bd/.style={circle, draw=black, inner sep=0pt, fill=black, minimum size=2mm}}
\tikzset{wd/.style={circle, draw=black, inner sep=0pt, fill=white, minimum size=2mm}}
\tikzset{Dynkin/.style={circle, draw=black, inner sep=0pt, fill=white, minimum size=2mm}}
\tikzstyle{ligne}=[draw, very thick] 
\tikzstyle{gridline}=[draw, gray] 
\tikzset{gauge/.style={circle, draw,inner sep=2.5pt}}
\tikzset{gaugeo/.style={circle, draw,inner sep=2.5pt,fill=orange}}
\tikzset{gauger/.style={circle, draw,inner sep=2.5pt,fill=red}}
\tikzset{gaugeb/.style={circle, draw,inner sep=2.5pt,fill=blue}}
\tikzset{gaugeg/.style={circle, draw,inner sep=2.5pt,fill=green}}
\tikzset{gaugegoodgreen/.style={circle, draw,inner sep=2.5pt,fill=goodgreen}}
\tikzset{gaugem/.style={circle, draw,inner sep=2.5pt,fill=magenta}}
\tikzset{hasse/.style={circle, fill,inner sep=2pt}}
\tikzset{d2/.style={circle, fill,inner sep=1.3pt}}
\tikzset{shrinky/.style={circle, fill,inner sep=1pt}}
\tikzset{sized/.style={circle, draw, inner sep=1.5pt}}
\tikzset{seven/.style={circle, draw,inner sep=3pt}}

\newcommand{\hsC}[1]{\hs\left[\mathcal C\left(\text{\Quiver{#1}}\right)\right]}
\newcommand{\hsCLST}[3]{\hs\left[\mathcal C\left(\text{\qquiver[#1]{#2}{#3}} \right)\right]}

\makeatletter
\DeclareRobustCommand{\rvdots}{%
  \vbox{
    \baselineskip4\p@\lineskiplimit\z@
    \kern-\p@
    \hbox{.}\hbox{.}\hbox{.}
  }}
\makeatother
\newcommand*{\qquiver}[3][]{$\mathcal{#2}^{#3}_{\ref{#1}}$}

\newcommand{\Figref}[1]{Figure~\ref{#1}}
\newcommand{\Quiver}[1]{$\mathcal Q_{\ref{#1}}$}
\newcommand{\surm}{\mathrm{SU}}

\newcommand{\urm}{\mathrm{U}}
\newcommand{\sorm}{\mathrm{SO}}
\newcommand{\orm}{\mathrm{O}}
\newcommand{\sprm}{\mathrm{Sp}}
\newcommand{\hs}{\mathrm{HS}}

\begin{document}

\title{Symmetry Mitosis and Hasse Diagram Diamonds:\\ A Note on Brane Configurations with $\mathrm{ON}^{0}$ Planes}

% Report number
\vspace*{-3cm} 
\begin{flushright}
{\tt DESY-25-038}\\
{\tt Imperial-TP-26-AH-01}\\
\end{flushright}
\author{Sam Bennett}
\email[\texttt{samuel.bennett18@imperial.ac.uk}]{}
\affiliation{Abdus Salam Centre for Theoretical Physics, Imperial College London, Prince Consort Road, SW7 2AZ, UK}

\author{Amihay Hanany}
\email[\texttt{a.hanany@imperial.ac.uk}]{}
\affiliation{Abdus Salam Centre for Theoretical Physics, Imperial College London, Prince Consort Road, SW7 2AZ, UK}

\author{Guhesh Kumaran}
\email[\texttt{guhesh.kumaran18@imperial.ac.uk}]{}
\affiliation{Abdus Salam Centre for Theoretical Physics, Imperial College London, Prince Consort Road, SW7 2AZ, UK}

\author{Lorenzo Mansi}
\email[\texttt{lorenzo.mansi@desy.de}]{}
\affiliation{Deutsches Elektronen-Synchrotron DESY, Notkestr.~85, 22607 Hamburg, Germany}

\begin{abstract}
\noindent This letter considers 3d $\mathcal{N}=4$ (unitary-)orthosymplectic quiver gauge theories originating from Type IIA and Type IIB brane systems with $\mathrm{ON}^0$ planes. Such theories lie outside the scope of present combinatorial techniques for Coulomb branch symmetry and symplectic stratification. It turns out that the correct prescription involves `symmetry mitosis': a common subset of nodes in two linear balanced chains source \emph{two} factors of a Coulomb branch global symmetry instead of one; the correct Coulomb branch Hasse diagram is obtained by a `doubling' procedure on that computed by naive quiver subtraction. Input from 6d SQFTs and little string theories allows for the construction of various `mitotic' magnetic quivers. The full Higgs branch Hasse diagram of minimal $(E_6,E_6)$ conformal matter is given. Additionally, a new Type I$'$ brane system using eight full D8 branes, negatively charged D6 branes, and $\mathrm{ON}^0$ planes is found corresponding to a product of $\mathrm{Spin}(32)$ instantons on $\mathbb C^2$. The corresponding 6d theory uses $\sprm(-1)$ gauge nodes which have the interpretation of bi-spinor matter of $\orm(a)$ and $\orm(12-a)$ for $a=0,1,\cdots,12$.
\end{abstract}

\maketitle

\section{Introduction}
The study of three-dimensional quiver gauge theories with eight real supercharges has proven particularly fruitful over the past few years. Following the characterisation of their quantum corrected Coulomb branch moduli space as a variety of dressed monopole operators, a cavalcade of results have accumulated concerning the Higgs branches of various higher-dimensional theories at finite and infinite coupling \cite{Sperling:2021fcf,Cabrera:2019dob,Bourget:2019rtl,Cabrera:2019izd,Mansi:2023faa,Bourget:2023dkj,Hanany:2022itc,Bourget:2021jwo,Bourget:2020mez,Bourget:2020gzi,Akhond:2024nyr,DeMarco:2025ugw,Bourget:2019aer}.
Broadly, such calculations rely on the interpretation of the Higgs branch of a $d=4/5/6$ dimensional theory as a moduli space of dressed monopole operators associated with an auxiliary object termed the \textit{magnetic quiver}. This moduli space also finds a realisation in terms of the Coulomb branch of a 3d $\mathcal{N}=4$ theory. Although common usage tends to collapse the distinction between this 3d theory and the magnetic quiver, it should be noted that the 3d theory does not generally correspond to the mirror symmetry result of the higher-dimensional theory's 3d compactification. As such, magnetic quivers are best thought of as a combinatorial tool for the construction of a moduli space, as opposed to a 3d $\mathcal{N}=4$ cousin of the original `electric' theory.

Crucially, the magnetic quiver incorporates quantum corrections to the Higgs branch of a higher-dimensional theory, such as Argyres-Douglas points in four dimensions, massless instantons in five dimensions, and tensionless strings in six dimensions. Such corrections often place the Higgs branch stratification of a $d\geq4$ dimensional theory beyond the reach of direct computation -- the magnetic quiver is in this case the only method at present able to recover the same information. 

The stratification of a symplectic singularity is naturally expressed using a Hasse diagram; from a physics perspective, this encodes various phases of a theory defined by the number of massless fields \cite{Bourget:2019aer} (in the case that the theory is Lagrangian) in its spectrum. In terms of the original higher-dimensional SQFT, the Hasse diagram captures the action of the generalised partial Higgs mechanism on the gauge group.\footnote{`Generalised' here refers to the fact that Higgs mechanism calculations are less obvious for non-Lagrangian theories.}

Despite its long gestation, much of the machinery developed for studying Coulomb branches of 3d $\mathcal{N}=4$ quiver gauge theories -- and thus magnetic quivers -- is still undergoing refinement \cite{Lawrie:2024wan,Gledhill:2021cbe}. New families of 3d $\mathcal{N}=4$ theories bring with them surprising combinatorics that require patient explication; techniques such as decoration \cite{Bourget:2019rtl,Bourget:2022ehw}, non-simple lacing \cite{Hanany2020UngaugingTheories,Grimminger:2024doq,Grimminger:2025fgj}, and orthosymplectic decay and fission \cite{Lawrie:2024wan} find their origins in precisely this context. This note further augments the combinatorics of magnetic quivers by considering the phenomenon of `symmetry mitosis', specifically in the context of unframed star-shaped orthosymplectic 3d $\mathcal{N}=4$ theories.

Although quiver subtraction techniques have in recent years helped uncover important information about the Higgs branches of various SQFTs, the algorithm remains a work-in-progress. Experience teaches that (unitary)-orthosymplectic quivers contain many complications absent from their unitary cousins. Despite recent attempts \cite{Lawrie:2024zon,Lawrie:2024wan} at delineating a generic prescription for quiver subtraction on unframed orthosymplectic quivers, various phenomena remain unknown to the literature. 

With this in mind, this short note aims to further plenish the toolbox of quiver subtraction techniques on 3d $\mathcal{N}=4$ theories with an aim to obtaining new Higgs branch phase diagrams for a range of six-dimensional SQFTs. Particular attention is paid to six-dimensional theories arising from brane systems containing an $\mathrm{ON}^0$ orientifold plane \cite{Kapustin:1998fa,Hanany:1999jy,Sen:1998ii,Hanany:2000fq}, which are associated with the presence of at least two copies of the same factor $F$ in the global symmetry of the magnetic quiver theory. Importantly, this breaks the magnetic balance rules in their current formulation \cite{Gledhill:2021cbe,Bennett:2024llh,Sperling:2021fcf}.

This class of examples arises in magnetic quiver constructions involving Type IIA brane systems consisting of D6-/D8- and NS5-branes with $\mathrm{O6}^-$, $\mathrm{O8}^-$ and $\mathrm{ON}^0$ orientifolds \cite{Hanany:1997gh, Cabrera:2019izd, Cabrera:2019dob}. These configurations support 6d SQFTs on the D6-brane worldvolume, whose Higgs branches/chiral rings are of particular interest. The usual understanding of these six-dimensional configurations and their corresponding Higgs branch Hasse diagrams relies on an alternative construction as compactifications of F-theory on an elliptically fibered Calabi--Yau threefold \cite{Morrison:1996na,Morrison:1996pp}. In this language, the geometry of the complex curves populating the base space of the Calabi--Yau determines the class of SQFTs under investigation, while the specifics of the fibration give the gauge group of the theory and its matter content. The global symmetry is encapsulated in non-compact curves in the base space, and the extraction of the pattern of partial Higgsing lies in an intricate analysis of the space of complex structure deformations of the threefold \cite{Bao:2024eoq,Bao:2025pxe}. In comparison, the quiver subtraction algorithm on magnetic quivers provides the same foliation using a comparatively simple set of rules that readily admit an algorithmic implementation \cite{Bourget:2023dkj,Bourget:2024mgn}. Of course, this is not quite the Faustian bargain it appears -- in some circumstances the more general geometric approach detects the existence of Higgsing directions that are unseen by the Type IIA brane system construction and hence the magnetic quiver. However, finding an appropriate translation of these transitions into the magnetic quiver language is often straightforward, and simply corresponds to an augmentation of the general magnetic algorithm. These extra rules are usually associated with multiplicity of various factors of the global symmetry.

Acquiring control over the cases where the usual balance rule \cite{Gledhill:2021cbe} fails (and the resulting Hasse diagram is incorrect) leads to the notion of `symmetry mitosis', studied explicitly here for the first time. By tracing copies of global symmetry factors to monopole contributions from a common subset of nodes in the quiver, the mitosis operation sheds light on the rules underlying the construction of Hasse diagrams, with an emphasis on their interpretation as distinguishable partial Higgsing directions.

The paper is thus structured as follows: Section~\ref{sec:Mitotic_Unitary}, aims to exploit 3d Type IIB D3-D5-NS5 brane constructions with $\mathrm{ON}^0$ planes leading to unitary-symplectic quiver gauge theories, making first contact with the concept of mitosis. These are intimately associated with product theories previously studied in the literature \cite{Sperling:2021fcf}. In Section~\ref{sec:Mitotic6dSQFT}, the focus is turned to the magnetic quiver of 6d SQFTs that exhibit mitosis. New quivers corresponding to Higgsed phases of Orbi-Instanton theories \cite{DelZotto:2014hpa} appear, and an interpretation of mitosis as a discrete symmetry in the F-theory compactification curve configuration in the base space of the Calai--Yau is given. Further extensions of this interpretation are in Section~\ref{sec:Mitotic6dLST}, where the magnetic quivers of six-dimensional little string theories are drawn. Finally, Section~\ref{sec:doubling} gives a more generic definition of mitosis, alongside an analysis of its effect on the Hasse Diagram.

\section{Symmetry Mitosis in Unitary Theories}\label{sec:Mitotic_Unitary}
The first indication that Type IIB brane systems involving $\mathrm{ON}^0$-planes have the potential to support 3d $\mathcal{N}=4$ product theories arose in \cite{Kapustin:1998fa} via the $D_2\cong A_1 \times A_1$ accidental isomorphism. 
This section considers sets of boundary conditions partitioning $N$ D3-branes ending on the orientifold into two sets of $N_1$ and $N_2$ branes satisfying $N=N_1+N_2$. The $\urm(N)$ gauge theory living on their world-volume is accordingly broken to $\urm(N_1)\times \urm(N_2)$. Terminating their other free end on an NS5-brane freezes the hypermultiplet between the two stacks. Generically, D5-branes contribute equally to the flavour symmetry of each gauge group although bound states with the orientifold can also be formed, leading to a double contribution only towards one of the two gauge nodes \cite{Kapustin:1998fa, Hanany:1999sj}. Depicting D3-branes as horizontal lines, NS5-branes as crossed circles, D5-branes as vertical lines, and the $\mathrm{ON^-}$-plane as an empty circle, the brane system just described is explicitly given as,
\begin{equation} \label{eqn:electric-branes+quiver2U(N)}
\begin{gathered}
  \begin{tikzpicture}
    %O5-
\ns{-4,0}{ns0}
\node[circle, draw, fill=white] (on) at (0,0) {};

\node[circle, draw, red, fill=white] (pairedns) at (-2,0) {};
	\node[cross out,red, draw] at (-2,0) () {};
    %D5
    \draw[thick] (-1.5,1.5)--(-1.5,-1.5);
    \node at (-1.1,1.0) () {\footnotesize $\cdots$};
    \draw[thick] (-0.75,1.5)--(-0.75,-1.5);
    \draw[decorate,decoration={brace, amplitude=3pt, raise=0.5ex}] (-1.5,1.5)--(-0.75,1.5) node[midway,label=above:{$h$}] () {};

    \draw[thick] (-3.6,1.5)--(-3.6,-1.5);
    \node at (-3.2,1.0) () {\footnotesize $\cdots$};
    \draw[thick] (-2.9,1.5)--(-2.9,-1.5);
    \draw[decorate,decoration={brace, amplitude=3pt, raise=0.5ex}] (-3.6,1.5)--(-2.9,1.5) node[midway,label=above:{$k$}] () {};
    
    \draw (ns0.north east)--(pairedns.north west) node[label={[xshift=-10pt,yshift=-7pt] $N_1$}] () {};
    \draw (ns0.south east)--(pairedns.south west) node[label={[xshift=-10pt,yshift=-20pt] $N_2$}] () {};
    \draw (pairedns.north east)--(on.north west) node[label={[xshift=-10pt,yshift=-7pt] $N_2$}] () {};
    \draw (pairedns.south east)--(on.south west) node[label={[xshift=-10pt,yshift=-20pt] $N_2$}] () {};
    \end{tikzpicture}  
\end{gathered} \, \longrightarrow\begin{gathered}
\begin{tikzpicture}
 \node[gauge,label=below:{$N_1$}] at (0,1.5) (1) {};
 \node[flavour,label=below:{$k$}] at (1.5,1.5) (2) {};
 \node[gauge,label=below:{$N_2$}] at (0,0) (3) {};
 \node[flavour,label=below:{$k+2h$}] at (1.5,0) (4) {};
 \node at (0.75,0.7) () {$\times$};
 \draw (1)--(2) (3)--(4);
\end{tikzpicture}
\end{gathered}
\end{equation}
where the NS5-brane coloured red arises from the $\text{ON}^{0}$ plane.

The magnetic quiver for the theory in \eqref{eqn:electric-branes+quiver2U(N)} has a computable Hilbert series\footnote{Generically, a 3d $\mathcal{N}=4$ theory has a Coulomb branch that typically consists of a singular part and possibly a smooth part. When the theory is ``good'' in the sense of \cite{Gaiotto:2008ak}, the Coulomb branch is a \emph{conical} symplectic singularity \cite{Beauville_2000}, and these are the spaces that can be studied with the monopole formula of \cite{Cremonesi:2013lqa}.} only for $h=0$, as shown in \cite{Sperling:2021fcf}, while restrictions to other cases have been discussed in \cite{Kapustin:1998fa, Ferlito:2016grh,Assel:2018exy, Sperling:2021fcf,Lawrie:2024zon}. The focus of this note lies in the case $N_1=N_2=N$, $k\ge2N$, $h=0$, for which the magnetic theory is the following unitary-orthosymplectic quiver
\begin{equation}\label{eqn:magnetic_quiver2U}
\begin{gathered}
    \resizebox{0.85\linewidth}{!}{\begin{tikzpicture}
        \node[gaugeb,label=below:{$2N$}] at (0,0) (usp) {};
        \node[flavourr,label=right:{$2$}] [above=0.7cm of usp] (f1) {};
        \node[gauge,label=below:{$2N$}] [left=0.8cm of usp] (u2n) {};
        \node [left=0.4cm of u2n] (dots) {\footnotesize $\cdots$};
        \node[gauge,label=below:{$2N$}] [left=0.4cm of dots] (u2nf) {};
        \node[flavour,label=right:{$1$}] [above=0.7cm of u2nf] (f2) {};
        \node[gauge,label=below:{$2N-1$}] [left=0.8cm of u2nf] (u2n-1) {};
        \node [left=0.4cm of u2n-1] (dots2) {\footnotesize $\cdots$};
        \node[gauge,label=below:{$2$}] [left=0.4cm of dots2] (u2) {};
        \node[gauge,label=below:{$1$}] [left=0.8cm of u2] (u1) {};

        \draw[decorate,decoration={brace, amplitude=3pt, raise=4.5ex}] (u2n.west)--(u2nf.east) node[midway,label={[yshift=-50pt]$k-2N$}] () {};

        \draw (f1)--(usp) (usp)--(u2n) (u2n)--(dots) (dots)--(u2nf) (u2nf)--(f2) (u2nf)--(u2n-1) (u2n-1)--(dots2) (dots2)--(u2) (u2)--(u1);
    \end{tikzpicture}
    }
\end{gathered}
\end{equation}
This 3d-mirror is found by moving to the phase where all D3 branes are suspended between D5 branes through a series of brane-creation transitions \cite{Hanany:1996ie}. An explicit Hilbert Series (HS) computation for the Coulomb branch of  {\Quiver{eqn:magnetic_quiver2U}} gives its global symmetry as $\surm(k) \times \surm(k)$, while on each branch the HS factorises as a product of HS of the mirror symmetric branch of \eqref{eqn:electric-branes+quiver2U(N)}, see \cite[Table 23]{Sperling:2021fcf}.
\begin{equation}
    \begin{gathered}
    \begin{tikzpicture}
    \node[gauge,label=below:{\footnotesize $N$}] at (0,0) (a) {};
    \node[gauge,label=below:{\footnotesize $N$}] at (1.5,0) (b) {};
    \node[gauge,label=above:{\footnotesize $N$}] at (0,1.5) (c) {};
    \node[gauge,label=above:{\footnotesize $N$}] at (1.5,1.5) (d) {};
    \node[flavour,label=left:{\footnotesize $k$}] at (-1,0) (a1) {};
    \node[flavour,label=left:{\footnotesize $k$}] at (-1,1.5) (c1) {};
    \node[flavour,label=right:{\footnotesize $k$}] at (2.5,0) (b1) {};
    \node[flavour,label=right:{\footnotesize $k$}] at (2.5,1.5) (d1) {};
    \draw[-] (a)--(b)--(c)--(d) (a)--(0.7,0.7) (0.8,0.8)--(d) (a1)--(a) (b1)--(b) (c1)--(c) (d1)--(d);
    \end{tikzpicture}
    \end{gathered}
\label{eqn:electric_A3_kapustin}
\end{equation}
Adding an extra $\mathrm{ON}^0$ plane to the configuration returns the affine $A_3\simeq D_3$ shaped quiver \eqref{eqn:electric_A3_kapustin} \cite{Kapustin:1998fa}. If instead the entire system is `doubled', so that there are now two NS5-branes with $n$ D3-branes between them and $\mathrm{ON}^{0}$ planes at each end, the resulting theory is $\mathcal{E}_{N,k}^{(n)}$, whose 3d mirror  $\mathcal{M}_{N,k}^{(n)}$ is the chain polymerisation \cite{Hanany:2024fqf, Lawrie:2023uiu} of two {\Quiver{eqn:magnetic_quiver2U}} quivers along a diagonal $\surm\left(n\right)\subseteq \surm(k) \times \surm(k)$ (for $n\leq 2N \leq k$) of the Coulomb branch isometry, i.e. a gauging of the common flavour symmetry on the electric side.
\begin{equation}\label{eqn:electric_polyUN}\mathcal{E}_{N,k}^{(n)}:=
    \begin{gathered}
        \begin{tikzpicture}
            \node[gauge,label=below:{\footnotesize $n$}] at (0,0) (c) {};
            \node[gauge,label=below:{\footnotesize $N$}] [above right=0.6cm and 0.6cm of c] (ar) {};
            \node[gauge,label=below:{\footnotesize $N$}] [below right=0.6cm and 0.6cm of c] (br) {};
            \node[gauge,label=below:{\footnotesize $N$}] [above left=0.6cm and 0.6cm of c] (al) {};
            \node[gauge,label=below:{\footnotesize $N$}] [below left=0.6cm and 0.6cm of c] (bl) {};

            \node[flavour,label=below:{\footnotesize $k-n$}] [right=0.8cm of ar] (far) {};
            \node[flavour,label=below:{\footnotesize $k-n$}] [right=0.8cm of br] (fbr) {};
            \node[flavour,label=below:{\footnotesize $k-n$}] [left=0.8cm of al] (fal) {};
            \node[flavour,label=below:{\footnotesize $k-n$}] [left=0.8cm of bl] (fbl) {};
            \draw (fbl)--(bl) (fal)--(al) (fbr)--(br) (far)--(ar) (bl)--(c) (al)--(c) (br)--(c) (ar)--(c);
        \end{tikzpicture}
    \end{gathered}
\end{equation}
In this case the theory is no longer a product. The polymerised mirror quiver $\mathcal{M}_{N,k}^{(n)}$ of $\mathcal{E}_{N,k}^{(n)}$ is:
\begin{equation}\label{eqn:magnetic_quiver4U}
\footnotesize\mathcal{M}_{N,k}^{(n)}:=
\begin{gathered}
\resizebox{0.75\columnwidth}{!}{
    \begin{tikzpicture}

        \node[gaugeb,label=below:{$2N$}] at (0,0) (usp) {};
        \node[flavourr,label=right:{$2$}] [above=0.7cm of usp] (f1) {};
        \node[gauge,label=below:{$2N$}] [left=0.8cm of usp] (u2n) {};
        \node [left=0.4cm of u2n] (dots) {\footnotesize $\cdots$};
        \node[gauge,label=below:{$2N$}] [left=0.4cm of dots] (u2nf) {};
        \node[flavour,label=right:{$1$}] [above=0.7cm of u2nf] (f2) {};
        \node[gauge,label=below:{$2N-1$}] [left=0.8cm of u2nf] (u2n-1) {};
        \node [left=0.4cm of u2n-1] (dots2) {\footnotesize $\cdots$};
        \node[circle] [left=0.4cm of dots2] (un) {};
        \node[circle] [above=1cm of un] (s1) {};
        \node[circle] [below=1cm of un] (s2) {};
        \node [left=0.4cm of un] (dots3) {\footnotesize $\cdots$};
        \node[gauge,label=below:{$2N-1$}] [left=0.4cm of dots3] (u2n-1sx) {};
        \node [left=0.4cm of u2n-1sx] (dots4) {\footnotesize $\cdots$};
        
        \draw[decorate,decoration={brace, amplitude=3pt, raise=4.5ex}] (u2n.west)--(u2nf.east) node[midway,label={[yshift=-50pt]$k-2N$}] () {};
        
     \draw[dashed,red] (s2)--(un.south) (un.north)--(s1);
             \node[gauge,label=below:{$n$}] [left=0.4cm of dots2] (un) {};
        
        \draw (f1)--(usp) (usp)--(u2n) (u2n)--(dots) (dots)--(u2nf) (u2nf)--(f2) (u2nf)--(u2n-1) (u2n-1)--(dots2) (dots2)--(un) (un)--(dots3) (dots3)--(u2n-1sx) (u2n-1sx)--(dots4);

    \end{tikzpicture}
    }
\end{gathered}
\end{equation}
where the vertical dashed line signals an axis of reflection. This proposal is supported via HS computations on a variety of triplets $(N,k,n)$, such as the case $(1,4,2)$ given in \eqref{HS:Nkn_141}.
\begin{equation}{\label{HS:Nkn_141}}
    \begin{aligned}
&\hs\left[\mathcal{C}\left(\mathcal{M}_{1,4}^{(2)}\right)\right]=\hs\left[\mathcal{H}\left(\mathcal{E}_{1,4}^{(2)}\right)\right]=\\
&= \frac{\left(\begin{aligned}1 &+ 10 t^2 + 93 t^4 + 
   550 t^6 + 2506 t^8 +  \\&+8694 t^{10}+ 23986 t^{12}+ 53476 t^{14} + 
   \\&+  98343 t^{16} +150588 t^{18} +193983 t^{20} +\\&+ 210864 t^{22}+ \cdots+
   t^{44}\end{aligned}\right)}{(1 - t^2)^5 (1 - t^4)^8 (1 - t^6)^3} \\
   \\
&\hs\left[\mathcal{C}\left(\mathcal{E}_{1,4}^{(2)}\right)\right]=\hs\left[\mathcal{H}\left(\mathcal{M}_{1,4}^{(2)}\right)\right]=\\
&= \frac{\left(\begin{aligned}1 &+ 3 t^2 + 22 t^4 + 67 t^6 + 244 t^8 + 541 t^{10} +\\&+  1231 t^{12} +
 2074 t^{14} + 3400 t^{16}  + \\&+ 4416 t^{18}+5542 t^{20} + 5646 t^{22}\cdots+ t^{44}\end{aligned}\right)}{(1 - t^2)^4 (1 - t^4)^3 (1 - t^6)^2 (1 -t^8)^3}
    \end{aligned}
\end{equation}

\section{Symmetry Mitosis and 6d SQFTs}\label{sec:Mitotic6dSQFT}
Brane systems in Type IIB consisting of D3-/D5-/NS5-branes with $\mathrm{O3}$, $\mathrm{O5}^-$, and $\mathrm{ON^0}$ orientifold planes \cite{Hanany:1999sj, Douglas:1996sw, Gimon:1996rq} extend the behaviour seen in \eqref{eqn:electric-branes+quiver2U(N)} to a class of (mostly) framed orthosymplectic quivers.\footnote{For examples, consult \cite{Sperling:2021fcf}.} Like in Section~\ref{sec:Mitotic_Unitary}, many of these setups correspond to quiver gauge theories which exhibit symmetry mitosis, i.e. the existence of more than one flavour symmetry read from different balanced nodes sharing a common subset. 

To probe unframed 3d $\mathcal{N}=4$ theories, it is helpful to consider magnetic quivers for electric theories arising in Type IIA brane systems instead. In particular, 6d $(1,0)$ SQFTs and little string theories (LSTs) can be constructed in Type IIA by engineering setups of D6-branes, NS5-branes, and flavour D8-branes, possibly in the presence of $\mathrm{O6}^-$, $\mathrm{O8}^-$, and $\mathrm{ON}^-$ orientifold planes \cite{Cabrera:2019dob, Cabrera:2019izd, DelZotto:2023myd, DelZotto:2022ohj, Lawrie:2023uiu, Lawrie:2024zon, Mansi:2023faa, Lawrie:2024wan, Sperling:2021fcf, Akhond:2024nyr, Bennett:2024llh}.\footnote{Vanishing Romans mass makes possible a lift of the Type IIA brane system to M-theory, leading to further insights.} Six-dimensional theories admit an alternative construction in F-theory as a compactification on an elliptically fibered Calabi--Yau threefold with intersecting complex curves populating the base space \cite{Morrison:1996na, Morrison:1996pp}.
Thus, 3d $\mathcal{N}=4$ orthosymplectic quiver gauge theories exhibiting symmetry mitosis can be studied as magnetic quivers for six-dimensional theories whose characteristics are known independently from F-theory constructions.\\

With this in mind, consider the class of 6d $(1,0)$ SQFTs known in the literature as Orbi-Instantons of type $G_{ADE}$ \cite{DelZotto:2014hpa}, where $G_{ADE}$ is a simply-laced Lie algebra. Discussion of LSTs will be deferred to Section \ref{sec:Mitotic6dLST}. Each Higgsed Orbi-Instanton theory is constructed from a homomorphism $\Gamma_{ADE}\rightarrow E_8$, where $\Gamma_{ADE}$ is a finite subgroup of $\surm(2)$, as well as an F-theory curve configuration. A Type IIA brane construction may also exist, although such is known only for $\Gamma_{ADE}=\{\mathbb{Z}_k,\mathbb{D}_{k+4}\}$ and the exceptional case $E_6$ \cite{Bennett:2024llh,Bao:2024wls}. The remaining binary-tetrahedron, -octahedron, and -icosahedron groups are associated with 6d theories comprising exceptional gauge algebras, whose construction as stacks of D$p$-branes on a non-perturbative orientifold plane is unknown \cite{Frey:2018vpw, Cabrera:2019izd, Lawrie:2023uiu}. For $\Gamma_{ADE}=\mathbb{D}_{k+4}$ the corresponding Type IIA D6-NS5-D8 brane construction contains $\mathrm{O6}^-$, $\mathrm{O8}^-$, and $\mathrm{ON}^0$ orientifold planes \cite{Hanany:1997gh}. Only cases that give rise to symmetry mitosis in the associated magnetic quiver are important for the purposes of this note. For the F-theory curve configuration associated with the 6d theory this is akin to demanding the existence of a discrete $S_n$ symmetry permuting the curves associated to the various flavours. The following curve configurations are among the simplest that give rise to a magnetic quiver with symmetry mitosis; for these theories, the discrete symmetry permuting the flavours is $S_2$. For \eqref{eqn:so8so8curveNK} this is the outer automorphism associated to the finite $D_{2N+2}$ Dynkin diagram and for \eqref{eqn:so4so4curve3}-\eqref{eqn:e6e6curve3} this is that for $D_3$.
\begin{equation}\label{eqn:so8so8curveNK}
\begin{aligned}
    &\mathrm{OI}_{N,k}:=\\
    &[\orm(4k+8)]\underbrace{\stackon{$1$}{$\mathfrak{sp}_{2k}$}\stackon{$4$}{$\mathfrak{o}_{8+4k}$} \cdots\stackon{$1$}{$\mathfrak{sp}_{2k}$}\stackon{\stackon{$4$}{$\mathfrak{o}_{8+4k}$}}{\stackon{\stackon{$1$}{$\mathfrak{sp}_k$}}{$[\orm(8)]$}}}_{N \text{ blocks }\stackon{\footnotesize$1$}{\footnotesize$\mathfrak{sp}_{2k}$}\stackon{\footnotesize$4$}{\footnotesize$\mathfrak{o}_{8+4k}$}} \stackon{$1$}{$\mathfrak{sp}_k$}[\orm(8)]
\end{aligned}
\end{equation}
\begin{equation}\label{eqn:so4so4curve3}
 \mathrm{OI}_{\orm(4)}:=[\sprm(5)\times \orm(1)] \stackon{\stackon{$3$}{$\mathfrak{o}_{12}$}}{\stackon{\stackon{$1$}{}}{$[\orm(4)]$}} 1\,[\orm(4)] 
\end{equation}
\begin{equation}\label{eqn:so5so5curve3}
 \mathrm{OI}_{\orm(5)}:=[\sprm(4)\times \orm(1)] \stackon{\stackon{$3$}{$\mathfrak{o}_{11}$}}{\stackon{\stackon{$1$}{}}{$[\orm(5)]$}} 1\,[\orm(5)] 
\end{equation}
\begin{equation}\label{eqn:so6so6curve3}
 \mathrm{OI}_{\orm(6)}:=[\sprm(3)\times \urm(1)] \stackon{\stackon{$3$}{$\mathfrak{o}_{10}$}}{\stackon{\stackon{$1$}{}}{$[\orm(6)]$}} 1\,[\orm(6)] 
\end{equation}
\begin{equation}\label{eqn:so7so7curve3}
 \mathrm{OI}_{\orm(7)}:=[\sprm(2)\times \sprm(1)] \stackon{\stackon{$3$}{$\mathfrak{o}_{9}$}}{\stackon{\stackon{$1$}{}}{$[\orm(7)]$}} 1\,[\orm(7)] 
\end{equation}
\begin{equation}\label{eqn:so8so8curve3}
 \mathrm{OI}_{\orm(8)}:=[\sprm(1)^3] \stackon{\stackon{$3$}{$\mathfrak{o}_{8}$}}{\stackon{\stackon{$1$}{}}{$[\orm(8)]$}} 1\,[\orm(8)] 
\end{equation}
\begin{equation}\label{eqn:so9so9curve3}
\mathrm{OI}_{\orm(9)}:=
 [\sprm(2)] \stackon{\stackon{$3$}{$\mathfrak{o}_{7}$}}{\stackon{\stackon{$1$}{}}{$[\orm(9)]$}} 1\,[\orm(9)] 
\end{equation}
\begin{equation}\label{eqn:f4f4curve3}
\mathrm{OI}_{F_4}:=
 [\sprm(1)] \stackon{\stackon{$3$}{$\mathfrak{g}_{2}$}}{\stackon{\stackon{$1$}{}}{$[F_4]$}} 1\,[F_4] 
\end{equation}
\begin{equation}\label{eqn:e6e6curve3}
\mathrm{OI}_{E_6}:=\stackon{\stackon{$3$}{$\mathfrak{su}_{3}$}}{\stackon{\stackon{$1$}{}}{$[E_6]$}} 1\,[E_6] 
\end{equation}

The curve notation of \eqref{eqn:so8so8curveNK}--\eqref{eqn:e6e6curve3} refers to the (negative) self-intersection number of the $\mathbb{CP}^1$s in the base space --- the reader is referred to the comprehensive review of \cite{Heckman:2018jxk} for more detailed explanations. Each of these six-dimensional theories has an electric Type IIA brane system obtained by taking the brane system of the conformal matter \cite{DelZotto:2014hpa} of type $\left(\orm(2k+8), \orm(2k+8)\right)$ \cite{Hanany:2022itc}, which include NS5 branes, D6 branes, and $\mathrm{O6}$, and then adding an $\mathrm{O8}^-$ plane with $8$ full D8-branes and an $\mathrm{ON}^0$ plane at intersection between the $\mathrm{O8}^-$ plane and the $\mathrm{O6}^-$. This operation corresponds to adding a $-1$ curve in the base space when considering the F-theory engineering.\\

The data in the curve configurations can be re-expressed as a 6d$\;\mathcal N=(1,0)$ quiver gauge theory. These are drawn below and are all reached through a Higgsing of the middle gauge node.\begin{align}
    &\begin{aligned}
        &Q\left(\mathrm{OI}_{\orm(4)}\right)=
        \begin{gathered}
            \begin{tikzpicture}
         \node[gaugeb] (spol) at (0,0) {};
        \node[align=center,anchor=north] (lab) at (spol.south) {$\sprm(0)$\\$-1$};
        \node[gauger] (sorm7) at (1,0) {};
        \node[align=center,anchor=north] (lab2) at (sorm7.south) {$\orm(12)$\\$-3$};
        \node[gaugeb] (spor) at (2,0) {};
        \node[align=center,anchor=north] (lab3) at (spor.south) {$\sprm(0)$\\$-1$};
        \node[flavourr, label=left:$\orm(4)$] (l) at (0,1) {};
        \node[flavourr, label=right:$\orm(4)$] (r) at (2,1){};
        \node[flavourb, label=above:$\sprm(5)$] (ml) at ({1-cos(62)},1) {};
        \node[flavourr, label=above:$\orm(1)$] (mr) at ({1+cos(62)},1) {};
        \draw[-] (l)--(spol)--(sorm7)--(spor)--(r);
        \draw[-] (ml)--(sorm7);
        \draw[-, color=red](mr)--(sorm7);
    \end{tikzpicture}
        \end{gathered}
    \end{aligned}\\
    &\begin{aligned}
        &Q\left(\mathrm{OI}_{\orm(5)}\right)=
        \begin{gathered}
            \begin{tikzpicture}
         \node[gaugeb] (spol) at (0,0) {};
        \node[align=center,anchor=north] (lab) at (spol.south) {$\sprm(0)$\\$-1$};
        \node[gauger] (sorm7) at (1,0) {};
        \node[align=center,anchor=north] (lab2) at (sorm7.south) {$\orm(11)$\\$-3$};
        \node[gaugeb] (spor) at (2,0) {};
        \node[align=center,anchor=north] (lab3) at (spor.south) {$\sprm(0)$\\$-1$};
        \node[flavourr, label=left:$\orm(5)$] (l) at (0,1) {};
        \node[flavourr, label=right:$\orm(5)$] (r) at (2,1){};
        \node[flavourb, label=above:$\sprm(4)$] (ml) at ({1-cos(62)},1) {};
        \node[flavourr, label=above:$\orm(1)$] (mr) at ({1+cos(62)},1) {};
        \draw[-] (l)--(spol)--(sorm7)--(spor)--(r);
        \draw[-] (ml)--(sorm7);
        \draw[-, color=red](mr)--(sorm7);
    \end{tikzpicture}
        \end{gathered}
    \end{aligned}\\
    &\begin{aligned}
        &Q\left(\mathrm{OI}_{\orm(6)}\right)=
        \begin{gathered}
            \begin{tikzpicture}
         \node[gaugeb] (spol) at (0,0) {};
        \node[align=center,anchor=north] (lab) at (spol.south) {$\sprm(0)$\\$-1$};
        \node[gauger] (sorm7) at (1,0) {};
        \node[align=center,anchor=north] (lab2) at (sorm7.south) {$\orm(10)$\\$-3$};
        \node[gaugeb] (spor) at (2,0) {};
        \node[align=center,anchor=north] (lab3) at (spor.south) {$\sprm(0)$\\$-1$};
        \node[flavourr, label=left:$\orm(6)$] (l) at (0,1) {};
        \node[flavourr, label=right:$\orm(6)$] (r) at (2,1){};
        \node[flavourb, label=above:$\sprm(3)$] (ml) at ({1-cos(62)},1) {};
        \node[flavourr, label=above:$\orm(2)$] (mr) at ({1+cos(62)},1) {};
        \draw[-] (l)--(spol)--(sorm7)--(spor)--(r);
        \draw[-] (ml)--(sorm7);
        \draw[-, color=red](mr)--(sorm7);
    \end{tikzpicture}
        \end{gathered}
    \end{aligned}\\
    &\begin{aligned}
        &Q\left(\mathrm{OI}_{\orm(7)}\right)=
        \begin{gathered}
            \begin{tikzpicture}
         \node[gaugeb] (spol) at (0,0) {};
        \node[align=center,anchor=north] (lab) at (spol.south) {$\sprm(0)$\\$-1$};
        \node[gauger] (sorm7) at (1,0) {};
        \node[align=center,anchor=north] (lab2) at (sorm7.south) {$\orm(9)$\\$-3$};
        \node[gaugeb] (spor) at (2,0) {};
        \node[align=center,anchor=north] (lab3) at (spor.south) {$\sprm(0)$\\$-1$};
        \node[flavourr, label=left:$\orm(7)$] (l) at (0,1) {};
        \node[flavourr, label=right:$\orm(7)$] (r) at (2,1){};
        \node[flavourb, label=above:$\sprm(2)$] (ml) at ({1-cos(62)},1) {};
        \node[flavourb, label=above:$\sprm(1)$] (mr) at ({1+cos(62)},1) {};
        \draw[-] (l)--(spol)--(sorm7)--(spor)--(r);
        \draw[-] (ml)--(sorm7);
        \draw[-, color=red](mr)--(sorm7);
    \end{tikzpicture}
        \end{gathered}
    \end{aligned}\\
    &\begin{aligned}
        &Q\left(\mathrm{OI}_{\orm(8)}\right)=\\
        &\qquad \qquad \begin{gathered}
            \begin{tikzpicture}
         \node[gaugeb] (spol) at (0,0) {};
        \node[align=center,anchor=north] (lab) at (spol.south) {$\sprm(0)$\\$-1$};
        \node[gauger] (sorm7) at (1,0) {};
        \node[align=center,anchor=north] (lab2) at (sorm7.south) {$\orm(8)$\\$-3$};
        \node[gaugeb] (spor) at (2,0) {};
        \node[align=center,anchor=north] (lab3) at (spor.south) {$\sprm(0)$\\$-1$};
        \node[flavourr, label=left:$\orm(8)$] (l) at (-0.5,0) {};
        \node[flavourr, label=right:$\orm(8)$] (r) at (2.5,0){};
        \node[flavourb, label=left:$\sprm(1)$] (ml) at ({1-cos(45)},{sin(45)}) {};
        \node[flavourb, label=above:$\sprm(1)$] (mm) at (1,1) {};
        \node[flavourb, label=right:$\sprm(1)$] (mr) at ({1+cos(45)},{sin(45)}) {};
        \draw[-] (l)--(spol)--(sorm7)--(spor)--(r);
        \draw[-, color=red] (ml)--(sorm7) node[pos=0.5,left]{$\textcolor{black}{S}$};
        \draw[-] (mm)--(sorm7) node[pos=0.5,right]{$V$};
        \draw[-, color=red](mr)--(sorm7) node[pos=0.5,right]{$\textcolor{black}{C}$};
    \end{tikzpicture}
        \end{gathered}
    \end{aligned}\\
    &\mathcal Q\left(\mathrm{OI}_{\orm(9)}\right)=\begin{gathered}
        \begin{tikzpicture}
         \node[gaugeb] (spol) at (0,0) {};
        \node[align=center,anchor=north] (lab) at (spol.south) {$\sprm(0)$\\$-1$};
        \node[gauger] (sorm7) at (1,0) {};
        \node[align=center,anchor=north] (lab2) at (sorm7.south) {$\orm(7)$\\$-3$};
        \node[gaugeb] (spor) at (2,0) {};
        \node[align=center,anchor=north] (lab3) at (spor.south) {$\sprm(0)$\\$-1$};
        \node[flavourr, label=left:$\orm(9)$] (l) at (0,1) {};
        \node[flavourr, label=right:$\orm(9)$] (r) at (2,1){};
        \node[flavourr, label=above:$\orm(5)$] (m) at (1,1) {};
        \draw[-] (l)--(spol)--(sorm7)--(spor)--(r);
        \draw[-,color=red] (m)--(sorm7);
    \end{tikzpicture}
    \end{gathered}\\
    &\mathcal Q\left(\mathrm{OI}_{F_4}\right)=\qquad \quad\begin{gathered}
        \begin{tikzpicture}
         \node[gaugeb] (spol) at (0,0) {};
        \node[align=center,anchor=north] (lab) at (spol.south) {$\sprm(0)$\\$-1$};
        \node[gaugeg] (sorm7) at (1,0) {};
        \node[align=center,anchor=north] (lab2) at (sorm7.south) {$G_2$\\$-3$};
        \node[gaugeb] (spor) at (2,0) {};
        \node[align=center,anchor=north] (lab3) at (spor.south) {$\sprm(0)$\\$-1$};
        \node[flavouro, label=left:$F_4$] (l) at (0,1) {};
        \node[flavouro, label=right:$F_4$] (r) at (2,1){};
        \node[flavourb, label=above:$\sprm(1)$] (m) at (1,1) {};
        \draw[-] (l)--(spol)--(sorm7)--(spor)--(r);
        \draw[-] (m)--(sorm7);
    \end{tikzpicture}
    \end{gathered}\\
    & \mathcal Q\left(\mathrm{OI}_{E_6}\right)=\qquad \quad\begin{gathered}
         \begin{tikzpicture}
         \node[gaugeb] (spol) at (0,0) {};
        \node[align=center,anchor=north] (lab) at (spol.south) {$\sprm(0)$\\$-1$};
        \node[gauge] (sorm7) at (1,0) {};
        \node[align=center,anchor=north] (lab2) at (sorm7.south) {$\surm(3)$\\$-3$};
        \node[gaugeb] (spor) at (2,0) {};
        \node[align=center,anchor=north] (lab3) at (spor.south) {$\sprm(0)$\\$-1$};
        \node[flavour,fill=violet!30, label=left:$E_6$] (l) at (0,1) {};
        \node[flavour, fill=violet!30, label=right:$E_6$] (r) at (2,1){};
        \draw[-] (l)--(spol)--(sorm7)--(spor)--(r);
    \end{tikzpicture}
     \end{gathered}
\end{align}where the labels $V,\;S,\;C$ refer to spinor, vector, and co-spinor representations of $\orm(8)$ respectively and the red line denotes a hypermultiplet in the bi-spinor representation. Notice that, unlike for $\orm(10)$, spinors of $\orm(11)$ and $\orm(12)$ are pseudoreal and hence allow for half-hypermultiplets.

Subsection \ref{subsectionA} analyses theories \eqref{eqn:so8so8curveNK}, \eqref{eqn:so9so9curve3}-\eqref{eqn:e6e6curve3} only, owing to the fact that the magnetic quivers for \eqref{eqn:so4so4curve3}-\eqref{eqn:so8so8curve3} have gauge nodes with negative balance, and so their Hilbert series are not computable.

Interestingly, the gauge nodes in the middle of \eqref{eqn:so9so9curve3}-\eqref{eqn:e6e6curve3} are $\surm(3)$, $G_2$ and $\sorm(7)$ -- this is the same sequence seen for the magnetic quivers in \cite{Bennett:2024llh}.

\subsection{$\mathrm{OI}_{N,k}$: An $\orm(8)\times \orm(8)$ Mitosis}
\label{subsectionA}

The curve configuration $\mathrm{OI}_{N,k}$ given in \eqref{eqn:so8so8curveNK} exhibits an explicit $S_2$ symmetry on the forked head, exchanging the two curves, each supporting an $\orm(8)$ flavour symmetry. The brane system and the magnetic quiver \qquiver[quiver:magneticOINK]{OI}{N,K} (all magnetic quivers for Orbi-Instanton SQFTs will be denoted by curly letters and corresponding equation number) for this theory were derived in \cite{Sperling:2021fcf,Lawrie:2024wan}, explicitly:
\begin{equation}\label{quiver:magneticOINK}   
\begin{gathered}
    \includegraphics[page=2,width=0.85\linewidth]{HSDoubling_Figures.pdf}
\end{gathered}
\end{equation}
The $\orm(8)$ symmetry undergoing mitosis appears in \qquiver[quiver:magneticOINK]{OI}{N, K} as two balanced chains, each starting from the $C_{2k+4+N}$ node and terminating at either of the $C$-nodes on the forked head of the quiver.
Interesting examples of this family of theories are:
\begin{itemize}
    \item[--] $N=0,k=0$ \cite{Sperling:2021fcf}, there are two effects at play in this case. Firstly, as $N=0$ the curve configuration $\mathrm{OI}_{0,0}$ reduces to two disconnected $(-1)$-curves each supporting $\orm(16)$ flavour symmetry, giving $\orm(16)\times\orm(16)$. This $\orm(16)$ flavour symmetry arises from the $\orm(8)$ on the left of $\mathrm{OI}_{0,0}$ combining with the two $\orm(8)$ flavour symmetries on the forked head. Secondly, because $k=0$ the fermionic zero modes that appear at infinite coupling enhance the symmetry to $E_8\times E_8$. The resulting orthosymplectic quiver's Coulomb Branch is a product of the closure of two minimal $E_8$ nilpotent orbits \cite[Table 3 and Table 9]{Sperling:2021fcf}. Equivalently, this moduli space may be thought of as the moduli space of two copies of one $E_8$ instanton on $\mathbb C^2$.
    
    \item[--] $N=0, k>0$ \cite{Sperling:2021fcf}, as in the case of $N=0, k=0$ the curve configuration $\mathrm{OI}_{0,k}$ reduces to two disconnected $(-1)$-curves each supporting an $\orm(4k+16)$ flavour symmetry. This flavour symmetry once again arises from the $\orm(4k+8)$ combining with the two $\orm(8)$ flavour symmetry factors in the forked head. There are also fermionic zero modes at infinite coupling in this case, transforming in the spinor representation of $\orm(4k+16)$, but these do not cause an enhancement of the flavour symmetry. In the magnetic quiver, these fermionic zero modes correspond to dressed monopole operators that appear at R-charges $(k+2)/2$.
    \item[--] $N=1,k=0$ \cite{Lawrie:2024wan}, where the $S_2$ symmetry of $\eqref{eqn:so8so8curveNK}$ naïvely enhances to $S_3$, realising an $\orm(8)^{3}$ symmetry with an $S_3$ which is not explicit in the magnetic quiver. Interestingly, the magnetic theory exhibits a 2-mitosis contributing two $\orm(8)$s alongside a third from an independent set of balanced nodes.
\end{itemize}
%Include the electric quiver for pedagogy.
Crucially, the $\mathrm{OI}_{N,k}$ theory can be regarded as a progenitor of the other theories exhibiting mitosis, as Higgsing the $\orm(4k+8)$ flavour symmetry preserves the forked head. For low enough $(N,k)$ (e.g. $N=1,k=0$), Higgsing produces all the other theories in \eqref{eqn:so8so8curve3}-\eqref{eqn:e6e6curve3}. The Higgs branch Hasse diagram of the $\mathrm{OI}_{N,k}$ theory is computed using the Decay and Fission algorithm \cite{Lawrie:2024zon, Bourget:2023dkj, Bourget:2024mgn} on the magnetic quiver and is given in Figure~\ref{fig:HasseOIso8} for the $\mathrm{OI}_{1,0}$ SQFT. In the following sections, some of these descending theories and their respective symmetry mitoses are studied in greater detail.

The brane conventions and dictionary to 6d electric quivers for the Type IIA systems considered here are found in \cite{Hanany:2022itc}.

\subsection{$\mathrm{OI}_{\orm(9)}$: An $\orm(9)\times \orm(9)$ Mitosis}
The $\mathrm{OI}_{\orm(9)}$ theory arises as a Higgsing of the $\mathrm{OI}_{1,0}$ SQFT. The first magnetic quiver with computable Hilbert series that arises from the Higgsing is \qquiver[quiv:2SO9]{OI}{\orm(7)}: 
\begin{equation}
    \begin{gathered}
    \resizebox{0.85\linewidth}{!}{\begin{tikzpicture}
    \node[gaugeb, label=right:$C_6$] (c6t) {};
    \node[gauger, label=below:$D_{13}$] (d13) [below=of c6t]{};
    \node[gaugeb, label=below:$C_{11}$] (c11) [left=of d13]{};
    \node[gauger, label=below:$D_{10}$] (d10) [left=of c11]{};
    \node[gaugeb, label=below:$C_8$] (c8) [left=of d10]{};
    \node[gauger, label=below:$D_7$] (d7) [left=of c8]{};
    \node[gaugeb, label=below:$C_5$] (c5) [left=of d7]{};
    \node[gauger, label=below:$D_4$] (d4) [left=of c5]{};
    \node[gaugeb, label=below:$C_2$] (c2) [left=of d4]{};
    \node[gauger, label=below:$D_2$] (d2) [left=of c2]{};
    \node[gaugeb, label=below:$C_1$] (c1) [left=of d2]{};
    \node[gauger, label=below:$D_1$] (d1) [left=of c1]{};
    \node[gaugeb, label=below:$C_8$] (c8r) [right=of d13]{};
    \node[gauger, label=below:$D_4$] (d4r) [right=of c8r]{};
    \draw[-] (d1)--(c1)--(d2)--(c2)--(d4)--(c5)--(d7)--(c8)--(d10)--(c11)--(d13)--(c8r)--(d4r) (d13)--(c6t);
\end{tikzpicture}}
\end{gathered}
\label{quiv:2SO9}
\end{equation}
The Coulomb branch Hilbert series of \qquiver[quiv:2SO9]{OI}{\orm(9)} can be  computed using the monopole formula, which supports the expected $\orm(5) \times \orm(9) \times \orm(9)$ global symmetry:
\begin{equation}
    \hsC{quiv:2SO9}=1+82t^2+O(t^3)
\end{equation}
In this case, the $\orm(5)$ flavour symmetry is sourced from the balanced tail $D_1-C_1-D_2-$ of \qquiver[quiv:2SO9]{OI}{\orm(9)}. The $\orm(9)\times \orm(9)$ symmetry emerges from the two balanced sets of nodes beginning at the $D_4$ gauge node and terminating at either of the $C$-type nodes of the fork. The two factors of $\orm(9)$ are characteristic of symmetry mitosis.

The Type IIA brane system associated with this theory is displayed in its electric and magnetic phases below. The magnetic phase is \begin{equation}
\begin{gathered}
\resizebox{0.85\linewidth}{!}{\begin{tikzpicture}
     \def\x{1.5cm};
         %D8
         \draw[-] (-3,-\x)--(-3,\x);
         \draw[-] (-2,-\x)--(-2,\x);
         \draw[-] (-1,-\x)--(-1,\x);
         \draw[-] (0,-\x)--(0,\x);
         \draw[-] (1,-\x)--(1,\x);
         \draw[-] (2,-\x)--(2,\x);
         \draw[-] (3,-\x)--(3,\x);
         \draw[-] (4,-\x)--(4,\x);
         \draw[-] (5,-\x)--(5,\x);
         \draw[-] (5,-\x)--(5,\x);
         \draw[-] (6,-\x)--(6,\x);
         \draw[-] (7,-\x)--(7,\x);
         \draw[-] (8,-\x)--(8,\x);
         \draw[-] (9,-\x)--(9,\x);
         \draw[-] (10,-\x)--(10,\x);
         \draw[dashed] (11,-\x)--(11,\x) node[pos=1,above]{$\mathrm{O8}^-$};
         
         %NS5
         \ns{11,0.5};
         \ns{11,-0.5};
         %D6
         \draw[-] (-2,0.25)--(9,0.25);
         \draw[-] (-2,-0.25)--(9,-0.25);

         \draw[-] (9,0.5)--(10,0.5);
         \draw[-] (9,-0.5)--(10,-0.5);

         \draw[-] (9,1)--(11,1);
         \draw[-] (11,1)--(10,0.9);
         \draw[-] (9,-1)--(11,-1);
         \draw[-] (11,-1)--(10,-0.9);

         %Tilde minus
         \draw[-] (-3,0)--(-2,0) (-1,0)--(0,0) (1,0)--(2,0) (3,0)--(4,0) (5,0)--(6,0) (7,0)--(8,0) (9,0)--(10,0);

        %ON
	   \node[circle, draw, fill=white, label=right:$\mathrm{ON}^{-}$] at (11,0) {};

        \node[label=right:$\times 4$] at (11,0.5){};
        \node[label=right:$\times 4$] at (11,-0.5){};
        \node[label=above:$1$] at (-1.5,0.2){};
        \node[label=above:$1$] at (-0.5,0.2){};
        \node[label=above:$2$] at (0.5,0.2){};
        \node[label=above:$2$] at (1.5,0.2){};
         \node[label=above:$4$] at (2.5,0.2){};
         \node[label=above:$5$] at (3.5,0.2){};
         \node[label=above:$7$] at (4.5,0.2){};
         \node[label=above:$8$] at (5.5,0.2){};
         \node[label=above:$10$] at (6.5,0.2){};
         \node[label=above:$11$] at (7.5,0.2){};
         \node[label=above:$13$] at (8.5,0.2){};
         \node[label=above:$6$] at (9.5,0.2){};
         \node[label=above:$8$] at (10.5,0.8){};

\end{tikzpicture}}\label{eq:SO9InfMagneticBrane}    
\end{gathered}
\end{equation} As a brief aside, the decay of the $C_3$ gauge node of the magnetic quiver $\mathrm{OI}_{\orm(9)}$ to a $C_2$ gauge node can be seen in the magnetic phase of the brane system given in \label{eq:OISO5InfMagneticBrane} by the exchange of the two D8 branes at the end of the interval housing the $C_3$ gauge node.

The electric phase of the brane system \label{eq:SO9InfMagneticBrane} is \begin{equation}
    \raisebox{-0.5\height}
    {\begin{tikzpicture}
        \ns{0,0};
        \ns{2,0};
        \ns{4,0};
        \draw[-] (1,-1)--(1,1)node[pos=0,below]{$5$} (5,-1)--(5,1)node[pos=0,below]{$9$};
        \draw[dashed] (6,-1)--(6,1) node[pos=1,above]{$\mathrm{O8}^-$};
        \node[circle, draw, fill=white, label=right:$\mathrm{ON}^{0}$] at (6,0) 
        {};

        \draw[-,color=red] (0,0.17)--(2,0.17);
        \draw[-,color=red] (0,-0.17)--(2,-0.17);
        \draw[-] (2,0.17)--(4,0.17);
        \draw[-] (2,-0.17)--(4,-0.17);

        \node[label=above:$-1$] at (0.5,0){};
        \node[label=above:$3$] at (3,0){};
        \draw[-] (2+0.17,0)--(4-0.17,0);
        \draw[dotted] (0.17,0)--(1,0) (5,0)--(5.83,0);
        \draw[dash dot] (1,0)--(1.83,0) (4.17,0)--(5,0);        
    \end{tikzpicture}}\label{eq:SO9InfElectricBrane}
\end{equation} Both flavour symmetry factors $\orm(5)\simeq\sprm(2)$ and $
\orm(9)$ are easily identifiable.

It is also possible to consider this gauge theory at the following finite coupling limit. \begin{equation}
\begin{gathered}
   \begin{tikzpicture}
         \node[gaugeb] (spol) at (0,0) {};
        \node[align=center,anchor=north] (lab) at (spol.south) {$\sprm(0)$\\$-1$\\$=0$};
        \node[gauger] (sorm7) at (1,0) {};
        \node[align=center,anchor=north] (lab2) at (sorm7.south) {$\orm(7)$\\$-3$\\$\neq0$};
        \node[gaugeb] (spor) at (2,0) {};
        \node[align=center,anchor=north] (lab3) at (spor.south) {$\sprm(0)$\\$-1$\\$=0$};
        \node[flavourr, label=left:$\orm(9)$] (l) at (0,1) {};
        \node[flavourr, label=right:$\orm(9)$] (r) at (2,1){};
        \node[flavourr, label=above:$\orm(5)$] (m) at (1,1) {};
        \node[] (ghost) at (-1,0){};
        \node[align=center,anchor=north] (lab4) at (ghost.south){$\phantom{\orm(0)}$\\$\phantom{-1}$\\$1/g^2$};
        \draw[-] (l)--(spol)--(sorm7)--(spor)--(r);
        \draw[-,color=red] (m)--(sorm7);
    \end{tikzpicture}\label{eq:2SO9Finite} 
\end{gathered}
\end{equation}The magnetic phase of the corresponding Type IIA brane system is obtained from the brane system in \eqref{eq:SO9InfMagneticBrane} by taking an NS5-brane off the $\mathrm{O8}^-$ plane and into the D8-brane interval with zero cosmological constant. This is the reverse of a small-$E_8$ instanton transition, and results in the brane system below.
\begin{equation}
\begin{gathered}
   \resizebox{0.85\linewidth}{!}{\begin{tikzpicture}
     \def\x{1.5cm};
         %D8
         \draw[-] (-3,-\x)--(-3,\x);
         \draw[-] (-2,-\x)--(-2,\x);
         \draw[-] (-1,-\x)--(-1,\x);
         \draw[-] (0,-\x)--(0,\x);
         \draw[-] (1,-\x)--(1,\x);
         \draw[-] (2,-\x)--(2,\x);
         \draw[-] (3,-\x)--(3,\x);
         \draw[-] (4,-\x)--(4,\x);
         \draw[-] (5,-\x)--(5,\x);
         \draw[-] (5,-\x)--(5,\x);
         \draw[-] (6,-\x)--(6,\x);
         \draw[-] (7,-\x)--(7,\x);
         \draw[-] (8,-\x)--(8,\x);
         \draw[-] (9,-\x)--(9,\x);
         \draw[-] (10,-\x)--(10,\x);
         \draw[dashed] (11,-\x)--(11,\x) node[pos=1,above]{$\mathrm{O8}^-$};
         
         %NS5
         \ns{11,0.5};
         \ns{11,-0.5};
         \ns{2.5,1};
         \ns{2.5,-1};
         %D6
         \draw[-] (-2,0.25)--(9,0.25);
         \draw[-] (-2,-0.25)--(9,-0.25);

         \draw[-] (9,0.5)--(10,0.5);
         \draw[-] (9,-0.5)--(10,-0.5);

         \draw[-] (9,1)--(11,1);
         \draw[-] (11,1)--(10,0.9);
         \draw[-] (9,-1)--(11,-1);
         \draw[-] (11,-1)--(10,-0.9);

         %Tilde minus
         \draw[-] (-3,0)--(-2,0) (-1,0)--(0,0) (1,0)--(2,0) (3,0)--(4,0) (5,0)--(6,0) (7,0)--(8,0) (9,0)--(10,0);

        %ON
	   \node[circle, draw, fill=white, label=right:$\mathrm{ON}^{-}$] at (11,0) {};

        \node[label=right:$\times 3$] at (11,0.5){};
        \node[label=right:$\times 3$] at (11,-0.5){};
        \node[label=above:$1$] at (-1.5,0.2){};
        \node[label=above:$1$] at (-0.5,0.2){};
        \node[label=above:$2$] at (0.5,0.2){};
        \node[label=above:$2$] at (1.5,0.2){};
         \node[label=above:$4$] at (2.5,0.2){};
         \node[label=above:$4$] at (3.5,0.2){};
         \node[label=above:$5$] at (4.5,0.2){};
         \node[label=above:$5$] at (5.5,0.2){};
         \node[label=above:$6$] at (6.5,0.2){};
         \node[label=above:$6$] at (7.5,0.2){};
         \node[label=above:$7$] at (8.5,0.2){};
         \node[label=above:$3$] at (9.5,0.2){};
         \node[label=above:$4$] at (10.5,1){};

\end{tikzpicture}}\label{eq:SO9InfMagneticBrane}
\end{gathered}
\end{equation}The associated magnetic quiver is straightforwardly given below, with Coulomb branch HS calculated to quadratic order in \eqref{hs:2SO9Finite}.\begin{equation}
\begin{gathered}
    \resizebox{0.8\linewidth}{!}{\begin{tikzpicture}
        \node[gaugeb, label=left:$C_3$] (c6t) {};
        \node[gauger, label=below:$D_{7}$] (d13) [below=of c6t]{};
        \node[gaugeb, label=below:$C_{6}$] (c11) [left=of d13]{};
        \node[gauger, label=below:$D_{6}$] (d10) [left=of c11]{};
        \node[gaugeb, label=below:$C_5$] (c8) [left=of d10]{};
        \node[gauger, label=below:$D_5$] (d7) [left=of c8]{};
        \node[gaugeb, label=below:$C_4$] (c5) [left=of d7]{};
        \node[gauger, label=below:$D_4$] (d4) [left=of c5]{};
        \node[gaugeb, label=below:$C_2$] (c2) [left=of d4]{};
        \node[gauger, label=below:$D_2$] (d2) [left=of c2]{};
        \node[gaugeb, label=below:$C_1$] (c1) [left=of d2]{};
        \node[gauger, label=below:$D_1$] (d1) [left=of c1]{};
        \node[gaugeb, label=below:$C_4$] (c8r) [right=of d13]{};
        \node[gauger, label=below:$D_2$] (d4r) [right=of c8r]{};
        \node[gaugeb, label=right:$C_1$] (c1t) [above=of d4]{};

        \draw[-] (d1)--(c1)--(d2)--(c2)--(d4)--(c5)--(d7)--(c8)--(d10)--(c11)--(d13)--(c8r)--(d4r) (d13)--(c6t) (c1t)--(d4);
    \end{tikzpicture}}\label{quiv:2SO9Finite}
\end{gathered}
\end{equation}
\begin{equation}
     \hsC{quiv:2SO9Finite}=1+82t^2+O(t^3)
\label{hs:2SO9Finite}
\end{equation}
The quadratic term of \eqref{hs:2SO9Finite} motivates an $\orm(9)$ 2-mitosis. This again arises from the two balanced chains of gauge nodes starting at the $D_4$ node and ending on the two $C$-type gauge nodes of the fork. The $\orm(5)$ factor is seen in the $D_1-C_1-D_2-$ chain on the left of the quiver.
\subsection{$\mathrm{OI}_{E_6}$: An $E_6\times E_6$ Mitosis}
The last example in this section is the $\mathrm{OI}_{E_6}$ SQFT, with magnetic quiver \qquiver[quiv:E6E6Magnetic]{OI}{E_6}:
\begin{equation}
\begin{gathered}
\resizebox{0.85\linewidth}{!}{\begin{tikzpicture}
         \node[gaugeb, label=right:$C_6$] (c6t) {};
        \node[gauger, label=below:$D_{13}$] (d13) [below=of c6t]{};
        \node[gaugeb, label=below:$C_{11}$] (c11) [left=of d13]{};
        \node[gauger, label=below:$D_{10}$] (d10) [left=of c11]{};
        \node[gaugeb, label=below:$C_8$] (c8) [left=of d10]{};
        \node[gauger, label=below:$D_7$] (d7) [left=of c8]{};
        \node[gaugeb, label=below:$C_5$] (c5) [left=of d7]{};
        \node[gauger, label=below:$D_4$] (d4) [left=of c5]{};
        \node[gaugeb, label=below:$C_2$] (c2) [left=of d4]{};
        \node[gauger, label=below:$D_1$] (d1) [left=of c2]{};
        \node[gaugeb, label=below:$C_8$] (c8r) [right=of d13]{};
        \node[gauger, label=below:$D_4$] (d4r) [right=of c8r]{};

        \draw[-] (d1)--(c2)--(d4)--(c5)--(d7)--(c8)--(d10)--(c11)--(d13)--(c8r)--(d4r) (d13)--(c6t);
    \end{tikzpicture}}
        \label{quiv:E6E6Magnetic}    
\end{gathered}
\end{equation}
In this case, there exist two balanced chains beginning at the $C_2$ node in the tail and terminating at either of the $C$-nodes in the fork. In addition the $D_1$ gauge node sources another $\urm(1)$ factor to the global symmetry. Together, these source an $\orm(10) \times \urm(1) \times \orm(10)\times \urm(1)$ factor in the flavour symmetry. This is enhanced due to spinor matter contribution to $E_6 \times E_6$, as reflected by the HS in \eqref{eqn:HSforE6E6}, and provides an example of $E_6 \times E_6$ symmetry mitosis. The counting of the spinor matter contribution is as an additional $2^6=64$ from the $6$ D-type gauge nodes in the magnetic quiver.\\

The $\mathrm{OI}_{E_6}$ theory merits further comment. The F-theory curve configuration shown in \eqref{eqn:e6e6curve3} is isomorphic to that for the minimal conformal matter theory of Type $(E_6, E_6)$ \cite{DelZotto:2014hpa}. In fact, the quiver \qquiver[quiv:E6E6Magnetic]{OI}{E_6} first appeared in the context of quotient quiver subtraction (i.e. $\surm(3)$ gauging of two copies of one $E_8$ instanton on $\mathbb C^2$) \cite{Bennett:2024llh}. The Higgs branch chiral ring of the 6d SQFT was studied through a Hilbert series computation on the magnetic quiver \qquiver[quiv:E6E6Magnetic]{OI}{E_6}; the result of \cite{Bennett:2024llh} is reproduced below:
\begin{align}
    \hs\left[\mathcal C(\text{\qquiver[quiv:E6E6Magnetic]{OI}{E_6}})\right]=&1+156 t^2+13703 t^4\nonumber\\&+876875 t^6+ O(t^{8}) \label{HS:E6E6ConfMatter} \\ \label{eqn:HSforE6E6}
    \mathrm{PL}\left[\hs\left[\mathcal C(\text{\qquiver[quiv:E6E6Magnetic]{OI}{E_6}})\right]\right]&=156 t^2 + 1457 t^4 \nonumber\\
    &+ 4627 t^6+O(t^{8}) 
\end{align}
The Type IIA brane system associated with the magnetic quiver is drawn in the magnetic phase as
\begin{equation}
\begin{gathered}
    \resizebox{0.85\linewidth}{!}{
   \begin{tikzpicture}
     \def\x{1.5cm};
         %D8
         \draw[-] (-1,-\x)--(-1,\x);
         \draw[-] (0,-\x)--(0,\x);
         \draw[-] (1,-\x)--(1,\x);
         \draw[-] (2,-\x)--(2,\x);
         \draw[-] (3,-\x)--(3,\x);
         \draw[-] (4,-\x)--(4,\x);
         \draw[-] (5,-\x)--(5,\x);
         \draw[-] (5,-\x)--(5,\x);
         \draw[-] (6,-\x)--(6,\x);
         \draw[-] (7,-\x)--(7,\x);
         \draw[-] (8,-\x)--(8,\x);
         \draw[-] (9,-\x)--(9,\x);
         \draw[-] (10,-\x)--(10,\x);
         \draw[dashed] (11,-\x)--(11,\x) node[pos=1,above]{$\mathrm{O8}^-$};

         %NS5
         \ns{11,0.5};
         \ns{11,-0.5};
         %D6
         \draw[-] (0,0.25)--(10,0.25);
         \draw[-] (0,-0.25)--(10,-0.25);

         \draw[-] (9,1)--(11,1);
         \draw[-] (11,1)--(10,0.9);
         \draw[-] (9,-1)--(11,-1);
         \draw[-] (11,-1)--(10,-0.9);

         %Tilde minus
         \draw[-] (-1,0)--(0,0) (1,0)--(2,0) (3,0)--(4,0) (5,0)--(6,0) (7,0)--(8,0) (9,0)--(10,0);

        %ON
	   \node[circle, draw, fill=white, label=right:$\mathrm{ON}^{-}$] at (11,0) {};

        \node[label=right:$\times 4$] at (11,0.5){};
        \node[label=right:$\times 4$] at (11,-0.5){};
        \node[label=above:$1$] at (0.5,0.2){};
        \node[label=above:$2$] at (1.5,0.2){};
         \node[label=above:$4$] at (2.5,0.2){};
         \node[label=above:$5$] at (3.5,0.2){};
         \node[label=above:$7$] at (4.5,0.2){};
         \node[label=above:$8$] at (5.5,0.2){};
         \node[label=above:$10$] at (6.5,0.2){};
         \node[label=above:$11$] at (7.5,0.2){};
         \node[label=above:$13$] at (8.5,0.2){};
         \node[label=above:$6$] at (9.5,0.2){};
         \node[label=above:$8$] at (10.5,0.9){};

         % \node at (-1.5,-\x) {$-$};
         % \node at (-0.5,-\x) {$\widetilde-$};
         % \node at (0.5,-\x) {$-$};
         % \node at (1.5,-\x) {$\widetilde-$};
         % \node at (2.5,-\x) {$-$};
         % \node at (3.5,-\x) {$\widetilde-$};
         % \node at (4.5,-\x) {$-$};
         % \node at (5.5,-\x) {$\widetilde-$};
         % \node at (6.5,-\x) {$-$};
         % \node at (7.5,-\x) {$\widetilde-$};
         % \node at (8.5,-\x) {$-$};
         % \node at (9.5,-\x) {$\widetilde-$};
         % \node at (10.5,-\x) {$-$};
    \end{tikzpicture}}
\end{gathered}
    \end{equation}
and in the electric phase
     \begin{equation}\begin{gathered}
      \resizebox{0.85\linewidth}{!}{\begin{tikzpicture}
        \def\x{1cm};
        % D8
        \draw (1,-\x)--(1,\x);
        \node[label=above:$2$] at (1,\x) {};
        \draw (5,-\x)--(5,\x);
        \node[label=above:$10$] at (5,\x) {};
        % NS5
        \ns{0,0};
        \ns{2,0};
        \ns{4,0};
        %O8
        \draw[dashed] (6,-\x)--(6,\x) node[pos=1,above]{$\mathrm{O8}^-$};
         %ON
	   \node[circle, draw, fill=white, label=right:$\mathrm{ON}^{0}$] at (6,0) {};
        %D6
        \draw[-,red] (0,0.2)--(2,0.2);
        \node[label=above:$-2$] at (1.5,0.2){};
        \draw[-,red] (0,-0.2)--(2,-0.2);
        \draw[dotted] (0.2,0)--(1.8,0);

        \draw[-] (2,0.2)--(4,0.2);
        \node[label=above:$3$] at (3,0.2){};
        \draw[-] (2,-0.2)--(4,-0.2);

        \draw[dotted] (4.2,0)--(5.8,0);

        % \node[] at (-0.5,-\x){$-$};
        % \node[] at (0.5,-\x){$+$};
        % \node[] at (1.5,-\x){$+$};
        % \node[] at (3,-\x){$-$};
        % \node[] at (4.5,-\x){$+$};
        % \node[] at (5.5,-\x){$+$};
        % \node[] at (7,-\x){$-$};
    \end{tikzpicture}}   
     \end{gathered}
    \end{equation}

\begin{figure*}
    \includegraphics[width=0.9\textwidth,page=1]{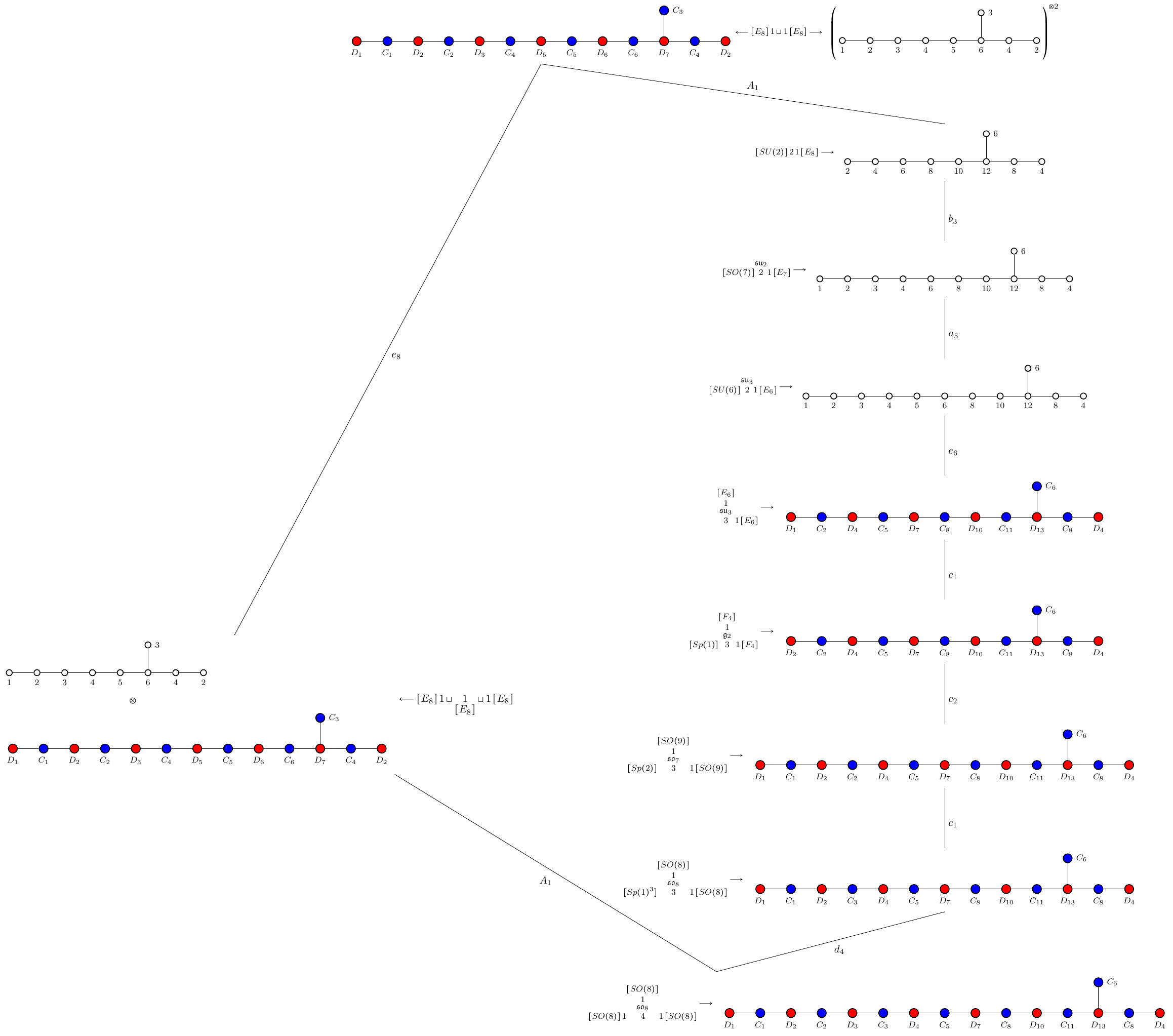}
    \caption{A portion of the Hasse Diagram for the magnetic quiver \qquiver[quiver:magneticOINK]{OI}{1, 0} based on the Decay and Fission algorithm. Next to each magnetic quiver is its corresponding electric 6d theory, which has been written in the F-theory curve notation. Notice that \Figref{fig:E6E6HasseClean} should also occur in Figure~\ref{fig:HasseOIso8} as a sub-diagram if the rest of the Higgsing directions were filled in.}
    \label{fig:HasseOIso8}
\end{figure*}

The claim of \cite{Bennett:2024llh} is further supported by the fact that \qquiver[quiv:E6E6Magnetic]{OI}{E_6}  naturally appears as a Higgsed phase of yet another mitotic quiver, the $\mathrm{OI}_{\orm(8)}$. As a decay product \cite{Lawrie:2024wan} of an orbi-instanton theory, the Hasse diagram of Figure~\ref{fig:HasseOIso8} precisely matches the F-theory-based Higgsing predictions of \cite{Bao:2024eoq, Bao:2025pxe}.

As before, the finite coupling limit of this theory may also be studied. Consider two $1/2$ M5 branes on a Klein $E_6$ singularity. In Type IIA language this corresponds to undoing a small $E_8$ instanton transition by bringing an NS5 brane off the $\mathrm{O8}^-$ plane. The electric quiver is 

\begin{equation}
\begin{gathered}
 \begin{tikzpicture}
         \node[gaugeb] (spol) at (0,0) {};
        \node[align=center,anchor=north] (lab) at (spol.south) {$\sprm(0)$\\$-1$\\$=0$};
        \node[gauge] (sorm7) at (1,0) {};
        \node[align=center,anchor=north] (lab2) at (sorm7.south) {$\surm(3)$\\$-3$\\$\neq0$};
        \node[gaugeb] (spor) at (2,0) {};
        \node[align=center,anchor=north] (lab3) at (spor.south) {$\sprm(0)$\\$-1$\\$=0$};
        \node[flavour, fill=violet!30,label=left:$E_6$] (l) at (0,1) {};
        \node[flavour, fill=violet!30,  label=right:$E_6$] (r) at (2,1){};
        \node[] (ghost) at (-1,0){};
        \node[align=center,anchor=north] (lab4) at (ghost.south){$\phantom{\orm(0)}$\\$\phantom{-1}$\\$1/g^2$};
        \draw[-] (l)--(spol)--(sorm7)--(spor)--(r);
    \end{tikzpicture}\label{eq:E6E6ElecFinite}   
\end{gathered}
\end{equation}with the corresponding magnetic theory\begin{equation}\begin{gathered}
  \resizebox{0.85\linewidth}{!}{\begin{tikzpicture}
         \node[gaugeb, label=left:$C_1$] (C1t) []{};
        \node[gauger, label=below:$D_4$] (D4l) [below=of C1t]{};
        \node[gaugeb, label=below:$C_2$] (C2l) [left=of D4l]{};
        \node[gauger, label=below:$D_1$] (D1l) [left=of C2l]{};
        \node[gaugeb, label=below:$C_4$] (C4l)[right=of D4l]{};
        \node[gauger, label=below:$D_5$] (D5l) [right=of C4l]{};
        \node[gaugeb, label=below:$C_5$] (C5l)[right=of D5l]{};
        \node[gauger, label=below:$D_6$] (D6l) [right=of C5l]{};
        \node[gaugeb, label=below:$C_6$] (C6l)[right=of D6l]{};
        \node[gauger, label=below:$D_7$] (D7) [right=of C6l]{};
        \node[gaugeb, label=below:$C_4$] (C4r) [right=of D7]{};
        \node[gauger, label=below:$D_2$] (D2r) [right=of C4r]{};
        \node[gaugeb, label=right:$C_3$] (C3t) [above=of D7]{};

        \draw[-] (D1l)--(C2l)--(D4l)--(C4l)--(D5l)--(C5l)--(D6l)--(C6l)--(D7)--(C4r)--(D2r) (D7)--(C3t) (D4l)--(C1t);
    \end{tikzpicture}}\label{eq:E6E6MagFinite}   
\end{gathered}
\end{equation}
Computing the Hilbert series associated to the magnetic quiver again demonstrates $E_6$ symmetry mitosis.
    \begin{align}
    \hs\left[\mathcal C(\text{\Quiver{eq:E6E6MagFinite}})\right]&=1 + 156 t^2 + 13859 t^4 \nonumber\\&+ 893669 t^6 + 45609733 t^8\nonumber\\&+1923636761t^{10} + O(t^{11})\\
    \mathrm{PL}\left[\hs\left[\mathcal C(\text{\Quiver{eq:E6E6MagFinite}})\right]\right]&=156 t^2 + 1613 t^4 - 2915 t^6 - 627017 t^8\nonumber\\&-1911458t^{10}+O(t^{11})
    \end{align}

\section{Symmetry Mitosis From 6d Little String Theories}\label{sec:Mitotic6dLST}
The previous section exploited brane constructions of 6d SQFTs that also arise in F-theory to generate mitotic, orthosymplectic magnetic 3d $\mathcal{N}=4$ quivers. A different class of 6d theories termed little string theories (LSTs) \cite{Seiberg:1997zk} may also enjoy doubling of global symmetry subgroups. LSTs are non-gravitational six-dimensional theories with an intrinsic string scale that cannot be set to zero. Interestingly, such a scale distinguishes them from both SQFTs and quantum gravity theories. The interested reader is referred to the extensive literature produced in recent years for a modern take on the subject \cite{DelZotto:2023nrb, DelZotto:2020sop, DelZotto:2014hpa, DelZotto:2022ohj, DelZotto:2022xrh, DelZotto:2023ahf, DelZotto:2023myd, Bhardwaj:2015oru, Ahmed:2023lhj, Ahmed:2024wve, Baume:2024oqn, Lawrie:2024zon, Lawrie:2023uiu, Mansi:2023faa}.

This section investigates LSTs whose magnetic quiver theories display symmetry mitosis by taking two (possibly Higgsed) orbi-instanton theories in six dimensions and gauging a subgroup of the flavour symmetry in such a way that the resulting theory is an LST \cite{DelZotto:2022ohj, Lawrie:2023uiu}. In the brane system, this simply corresponds to taking two $\mathrm{ON}^{0}$ planes instead of one, whose separation is precisely the scale of the resulting LST. \footnote{The gauging action corresponds to an addition of a compact curve in the base space configuration of the F-theory engineering such that the adjacency matrix of the configuration is semi-negative defined with only one zero eigenvalue \cite{Bhardwaj:2015oru}.} 

\subsection{$\mathrm{LST}_{N,k}$: A Double $\orm(8)\times \orm(8)$ Mitosis}
Considering two copies of the $\mathrm{OI}_{N,k}$ theory, it is possible to gauge the common $\orm(4k+8)$ flavour symmetry resulting in a little string theory, labelled $\mathrm{LST}_{N,k}$, whose curve configuration is:
\begin{equation}\label{eqn:lstso8so8curveNK}
\begin{aligned}
    &\mathrm{LST}_{N,k}:=\\
    &[\orm(8)]\stackon{$1$}{$\mathfrak{sp}_k$}\stackon{\stackon{$4$}{$\mathfrak{o}_{8+4k}$}}{\stackon{\stackon{$1$}{$\mathfrak{sp}_k$}}{$[\orm(8)]$}}\underbrace{\stackon{$1$}{$\mathfrak{sp}_{2k}$}\stackon{$4$}{$\mathfrak{o}_{8+4k}$} \cdots\stackon{$1$}{$\mathfrak{sp}_{2k}$}\stackon{\stackon{$4$}{$\mathfrak{o}_{8+4k}$}}{\stackon{\stackon{$1$}{$\mathfrak{sp}_k$}}{$[\orm(8)]$}}}_{N \text{ blocks }\stackon{\footnotesize$1$}{\footnotesize$\mathfrak{sp}_{2k}$}\stackon{\footnotesize$4$}{\footnotesize$\mathfrak{o}_{8+4k}$}} \stackon{$1$}{$\mathfrak{sp}_k$}[\orm(8)]
\end{aligned}
\end{equation}
where a relabelling of the $N$ is made for convenience. Notice that a further ${\stackon{$4$}{$\mathfrak{o}_{8+4k}$}}$ curve appears in the middle of the spine following the introduction of tensor and vector multiplets necessary for the above gauging. 

A brane construction for this theory in Type IIA is presented below, with explicit dependence on $N$ and $k$:\begin{equation}
    \begin{gathered}
        \resizebox{0.85\linewidth}{!}{\begin{tikzpicture}

        \ns{2,0};
        \ns{4,0};
        \ns{6,0};
        \ns{8,0};
        \ns{10,0};
        \ns{12,0};
        \ns{14,0};
        
        \draw[-] (1,-1)--(1,1)node[pos=0,below]{$8$} (15,-1)--(15,1)node[pos=0,below]{$8$};
        \draw[dashed] (16,-1)--(16,1) node[pos=1,above]{$\mathrm{O8}^-$};
        \draw[dashed] (0,-1)--(0,1) node[pos=1,above]{$\mathrm{O8}^-$};
        
        \node[circle, draw, fill=white, label=left:$\mathrm{ON}^{0}$] at (0,0) 
        {};

        \node[circle, draw, fill=white, label=right:$\mathrm{ON}^{0}$] at (16,0) 
        {};
        
        \draw[-] (0,0.17)--(8.7,0.17) (9.3,0.17)--(16,0.17);
        \draw[-] (0,-0.17)--(8.7,-0.17) (9.3,-0.17)--(16,-0.17);

        \node at (9,0){$\cdots$};

        \node[label=above:$2k+4$] at (3,0){};
        \node[label=above:$2k$] at (5,0){};
        \node[label=above:$2k+4$] at (7,0){};
        \node[label=above:$2k$] at (11,0){};
        \node[label=above:$2k+4$] at (13,0){};
        \node[label=above:$k$] at (1.5,0){};
        \node[label=above:$k$] at (14.5,0){};
        \node[label=below:$k$] at (1.5,0){};
        \node[label=below:$k$] at (14.5,0){};

        \draw [decorate, 
    decoration = {brace,
        raise=20pt,
        amplitude=5pt}] (2,0) --  (14,0) node[pos=0.5,above=25pt,black]{$2N+2$ half $\mathrm{NS}5$};
        
        \draw[dotted] (0.2,0)--(1.8,0) (4.2,0)--(5.8,0) (10.2,0)--(11.8,0) (14.2,0)--(15.8,0);
    \end{tikzpicture}}\label{eq:GeneralDQuiverBrane}
    \end{gathered}
\end{equation}
This brane system consists of two parallel $\mathrm{O8}^-$ planes with D8, D6, NS5, branes and $\mathrm{O6}$ and $\mathrm{ON}^0$ orientifolds. The two parallel $\mathrm{O8}^-$ planes is a Type I$'$ brane system. In the absence of $\mathrm{ON}^0$ planes there are $16$ full D8 branes in the interval, however the presence of the $\mathrm{ON}^0$ planes allows for only $8$ full D8 branes in the interval and not $16$. Nevertheless, the resulting gauge theory on the D6 branes is anomaly free. Due to the $\mathrm{ON}^0$ planes, the leftmost and rightmost $\mathrm{NS}5$ branes see the image of the $8$ half D8 branes through the orientifold plane in addition to those in the interval. Essentially, this double-counts the half D8 branes giving the correct cancellation of the anomaly.

The brane system admits an overall scale given by the separation of the $\mathrm{O8}^-$ planes which is the string scale in LSTs.

In the electric phase the brane system gives the following electric quiver \Quiver{eq:LSTNK} describing the theory on the D6 branes \begin{equation}
    \begin{gathered}
    \resizebox{\linewidth}{!}{
   \begin{tikzpicture}
         \node[gaugeb] (spkl) at ({1-sin(30)},{-cos(30)}){};
         \node[align=center, anchor=north] (labspkl) at (spkl.south){$\sprm(k)$\\$-1$};

         \node[gaugeb] (spklt) at ({1-sin(30)},{cos(30)}){};
         \node[align=center, anchor=south] (labskplt) at (spklt.north){$\sprm(k)$\\$-1$};

         \node[gauger] (ol) at (1,0){};
         \node[align=center, anchor=east] (labol) at (ol.west){$\orm(4k+8)$\\$-4$};

         \node[gaugeb] (sp2kl) at (2,0){};
         \node[align=center, anchor=north] (labsp2kl) at (sp2kl.south){$\sprm(2k)$\\$-1$};

          \node[gauger] (om) at (3,0){};
         \node[align=center, anchor=south] (labom) at (om.north){$\orm(4k+8)$\\$-4$};

         \node[] (cdots) at (4,0) {$\cdots$};

         \node[gaugeb] (sp2kr) at (5,0){};
         \node[align=center, anchor=north] (labsp2kr) at (sp2kr.south){$\sprm(2k)$\\$-1$};

          \node[gauger] (or) at (6,0){};
         \node[align=center, anchor=west] (labor) at (or.east){$\orm(4k+8)$\\$-4$};

         \node[gaugeb] (spkr) at ({6+sin(30)},{-cos(30)}){};
         \node[align=center, anchor=north] (labspkr) at (spkr.south){$\sprm(k)$\\$-1$};

         \node[gaugeb] (spkrt) at ({6+sin(30)},{cos(30)}){};
         \node[align=center, anchor=south] (labskprt) at (spkrt.north){$\sprm(k)$\\$-1$};
 
        \node[flavourr, label=left:$\orm(8)$] (fspkl) at ({-sin(30)},{-cos(30)}) {};
        \node[flavourr, label=left:$\orm(8)$] (fspklt) at ({-sin(30)},{cos(30)}){};

        \node[flavourr, label=right:$\orm(8)$] (fspkr) at ({7+sin(30)},{-cos(30)}) {};
        \node[flavourr, label=right:$\orm(8)$] (fspkrt) at ({7+sin(30)},{cos(30)}){};
        
        \draw[-] (fspkl)--(spkl)--(ol)--(sp2kl)--(om)--(cdots)--(sp2kr)--(or)--(spkr)--(fspkr) (fspklt)--(spklt)--(ol) (fspkrt)--(spkrt)--(or);

        \draw [decorate, 
    decoration = {brace,
        raise=30pt,
        amplitude=5pt}] (6,0) --  (2,0) node[pos=0.5,below=35pt,black]{$2N$ gauge nodes};      
    \end{tikzpicture}}\label{eq:LSTNK} 
\end{gathered}
\end{equation}this gauge theory has clear field theory interpretation when $k\geq0$ and $N\geq 0$. The meaning of $\sprm(0)$ (when $k=0$) in field theory is as a dynamical tensor multiplet.

Going to the magnetic phase of the brane system involves suspending all D6 branes between D8 branes through a sequence of Hanany-Witten transitions \cite{Hanany:1996ie}. The classical Higgs branch dimension of the 6d theory is $\mathrm{dim}\;\mathcal H_{\text{classical}}=30k$ which is the same as the dimension of $k$ $\mathrm{Spin}(32)/\mathbb{Z}_2$ instantons on $\mathbb C^2$, since the dual Coxeter number of $\mathrm{Spin}(32)/\mathbb Z_2$ is $30$. The magnetic quiver in this case is 
\begin{equation}\label{quiver:magneticLSTNKVeryFinite}
\begin{gathered}
\resizebox{\linewidth}{!}{\begin{tikzpicture}
\node[gauger, label=below:$B_{2k}$] (d4) at (0,0) {};
\node[gaugeb, label=above:$C_{2k}$] (cnp4) at (1,0) {};
\node[gauger, label=below:$B_{2k}$] (d2np5) at (2,0) {};
\node[gaugeb, label=above:$C_{2k}$] (c3np5) at (3,0) {};
\node[gauger, label=below:$B_{2k}$] (d4np6) at (4,0) {};
\node[gaugeb, label=above:$C_{2k}$] (c5np6) at (5,0) {};
\node[gauger, label=below:$B_{2k}$] (d6np7) at (6,0) {};
\node[gaugeb, label=above:$C_{k}$] (c4np4) at (7,0) {};
\node[gaugeb, label=right:$C_{k}$] (c3np3) at (6,1) {};

\node[flavourr, label=right:$B_0$] (bflavr) at (8,0){};

\node[flavourr, label=above:$B_0$] (bflavrt) at (6,2){};

\draw[-] (d4)--(cnp4)--(d2np5)--(c3np5)--(d4np6)--(c5np6)--(d6np7)--(c4np4)--(bflavr) (d6np7)--(c3np3)--(bflavrt);

\node[gaugeb, label=above:$C_{2k}$] (cmp4) at (-1,0) {};
\node[gauger, label=below:$B_{2k}$] (d2mp5) at (-2,0) {};
\node[gaugeb, label=above:$C_{2k}$] (c3mp5) at (-3,0) {};
\node[gauger, label=below:$B_{2k}$] (d4mp6) at (-4,0) {};
\node[gaugeb, label=above:$C_{2k}$] (c5mp6) at (-5,0) {};
\node[gauger, label=below:$B_{2k}$] (d6mp7) at (-6,0) {};
\node[gaugeb, label=above:$C_{k}$] (c4mp4) at (-7,0) {};
\node[gaugeb, label=left:$C_{k}$] (c3mp3) at (-6,1) {};

\node[flavourr, label=left:$B_0$] (bflavl) at (-8,0){};

\node[flavourr, label=above:$B_0$] (bflavlt) at (-6,2){};

\draw[-] (d4)--(cmp4)--(d2mp5)--(c3mp5)--(d4mp6)--(c5mp6)--(d6mp7)--(c4mp4)--(bflavl) (d6mp7)--(c3mp3)--(bflavlt);

\node[flavourb,label=above:$C_{N+1}$] (cflav) at (0,1){};

\draw[-] (cflav)--(d4);

\end{tikzpicture}}
\end{gathered}
% \begin{gathered}
%     \includegraphics[page=3,width=0.85\linewidth]{HSDoubling_Figures.pdf}
% \end{gathered}   
\end{equation}

It is possible for the half $\mathrm{NS}5$ branes to recombine on the $\mathrm{O6}$ plane and come off. The $\mathrm{ON}^0$ planes should be thought of as $\mathrm{ON}^-$ planes, each with a virtual $\mathrm{NS}5$, giving a total of $2N+4$ half $\mathrm{NS}5$ branes. There are hence $N+2$ full $\mathrm{NS}5$ branes in this brane system. There are two cases to consider when recombining $\mathrm{NS}5$ branes on the orientifold; namely when there are $2k+4$ or $2k$ D6 branes between them. In the first case the $\mathrm{NS}5$ branes recombine and come off as usual and in the magnetic quiver there is a $C_1$ gauge node. In the second case there are four additional frozen D6 branes and so when the $\mathrm{NS}5$ branes come off there are additional flavours associated to those frozen branes, this flavour could be of type B or D depending on the particular configuration \cite{Akhond:2024nyr}. When all of the D6 branes are suspended between D8 branes and all the $\mathrm{NS}5$ branes are separated and off the orientifold planes, the corresponding gauge theory has half of the gauge groups at finite coupling. The magnetic quiver in this case is

\begin{equation}\label{quiver:magneticLSTNKFinite}
\begin{gathered}
\resizebox{\linewidth}{!}{\begin{tikzpicture}
\node[gauger, label=below:$D_{2k+4}$] (d4) at (0,0) {};
\node[gaugeb, label=above:$C_{2k+3}$] (cnp4) at (1,0) {};
\node[gauger, label=below:$D_{2k+3}$] (d2np5) at (2,0) {};
\node[gaugeb, label=above:$C_{2k+2}$] (c3np5) at (3,0) {};
\node[gauger, label=below:$D_{2k+2}$] (d4np6) at (4,0) {};
\node[gaugeb, label=above:$C_{2k+1}$] (c5np6) at (5,0) {};
\node[gauger, label=below:$D_{2k+1}$] (d6np7) at (6,0) {};
\node[gaugeb, label=right:$C_{k}$] (c4np4) at (7,0) {};
\node[gaugeb, label=above:$C_{k}$] (c3np3) at (6,1) {};

\draw[-] (d4)--(cnp4)--(d2np5)--(c3np5)--(d4np6)--(c5np6)--(d6np7)--(c4np4) (d6np7)--(c3np3);

\node[gaugeb, label=above:$C_{2k+3}$] (cmp4) at (-1,0) {};
\node[gauger, label=below:$D_{2k+3}$] (d2mp5) at (-2,0) {};
\node[gaugeb, label=above:$C_{2k+2}$] (c3mp5) at (-3,0) {};
\node[gauger, label=below:$D_{2k+2}$] (d4mp6) at (-4,0) {};
\node[gaugeb, label=above:$C_{2k+1}$] (c5mp6) at (-5,0) {};
\node[gauger, label=below:$D_{2k+1}$] (d6mp7) at (-6,0) {};
\node[gaugeb, label=left:$C_{k}$] (c4mp4) at (-7,0) {};
\node[gaugeb, label=above:$C_{k}$] (c3mp3) at (-6,1) {};

\node[gaugeb, label=left:$C_1$] (c1l) at (-1,1){};
\node[gaugeb, label=right:$C_1$] (c1r) at (1,1){};

\node[] (cdots) at (0,1){$\cdots$};

\draw[-] (d4)--(cmp4)--(d2mp5)--(c3mp5)--(d4mp6)--(c5mp6)--(d6mp7)--(c4mp4) (d6mp7)--(c3mp3);

\draw[-] (c1l)--(d4)--(c1r);

\draw [decorate, 
    decoration = {brace,
        raise=10pt,
        amplitude=5pt}] (-1,1) --  (1,1) node[pos=0.5,above=15pt,black]{$N+2$};      
\end{tikzpicture}}
\end{gathered}
% \begin{gathered}
%     \includegraphics[page=3,width=0.85\linewidth]{HSDoubling_Figures.pdf}
% \end{gathered}   
\end{equation}where the $N+2$ $\mathrm{NS}5$ branes are represented by the bouquet of $N+2$ $C_1$ gauge nodes. The magnetic quiver contains an affine $D_{16}$ shape which descends from the Heterotic $\mathrm{Spin}(32)/\mathbb Z_2$ string.

The 6d theory has various phases corresponding to the values of the gauge couplings. In the brane system this is realised through the positions of the $\mathrm{NS}5$ branes. The $\mathrm{NS}5$ branes may either remain in their place, move onto the left or right $\mathrm{O8}^-$ planes, or the $\mathrm{NS}5$ branes may become coincident. The most generic magnetic quiver will be parametrised by the following data: $N$ (related to the number of $\mathrm{NS}5$ branes), $k$ (related to the number of D6 branes), $L$, $R$ which define how many $\mathrm{NS}5$ branes move onto the left and right $\mathrm{O8}^-$ respectively, and finally a partition $\{n_1,\cdots,n_j\}$ of $N+2-L-R$ which defines how many of the $\mathrm{NS}5$ branes in the middle of the interval become coincident. Moving an $\mathrm{NS}5$ brane onto an $\mathrm{O8}^-$ plane corresponds to setting the gauge coupling of an $\sprm(k)$ or $\sprm(2k)$ gauge node in the 6d theory to infinity, there are $N+2$ gauge nodes of this type corresponding to the $N+2$ $\mathrm{NS}5$ branes. Making $\mathrm{NS}5$ branes coincident corresponds to setting the gauge coupling of the $\orm(4k+8)$ gauge groups to infinity, there are $N+1$ gauge nodes of this type corresponding to the $N+1$ intervals between the $N+2$ $\mathrm{NS}5$ branes. Note that not all of the gauge couplings in the 6d theory cannot be set to infinity simultaneously, the lingering scale in the theory is a feature of LSTs. 

The most generic magnetic quiver for this scenario is drawn below \begin{equation}\label{quiver:magneticLSTNKGeneric}
\begin{gathered}
\resizebox{\linewidth}{!}{\begin{tikzpicture}
\node[gauger, label=below:$D_{2k+4}$] (d4) at (0,0) {};
\node[gaugeb, label=above:$C_{2k+R+3}$] (cRp4) at (1,0) {};
\node[gauger, label=below:$D_{2k+2R+3}$] (d2Rp5) at (2,0) {};
\node[gaugeb, label=above:$C_{2k+3R+2}$] (c3Rp5) at (3,0) {};
\node[gauger, label=below:$D_{2k+4R+2}$] (d4Rp6) at (4,0) {};
\node[gaugeb, label=above:$C_{2k+5R+1}$] (c5Rp6) at (5,0) {};
\node[gauger, label=below:$D_{2k+6R+1}$] (d6Rp7) at (6,0) {};
\node[gaugeb, label=above:$C_{k+4R}$] (c4Rp4) at (7,0) {};
\node[gauger, label=below:$D_{2R}$] (d2Rp2) at (8,0) {};
\node[gaugeb, label=above:$C_{k+3R}$] (c3Rp3) at (6,1) {};

\draw[-] (d4)--(cRp4)--(d2Rp5)--(c3Rp5)--(d4Rp6)--(c5Rp6)--(d6Rp7)--(c4Rp4)--(d2Rp2) (d6Rp7)--(c3Rp3);

\node[gaugeb, label=above:$C_{2k+L+3}$] (cLp4) at (-1,0) {};
\node[gauger, label=below:$D_{2k+2L+3}$] (d2Lp5) at (-2,0) {};
\node[gaugeb, label=above:$C_{2k+3L+2}$] (c3Lp5) at (-3,0) {};
\node[gauger, label=below:$D_{2k+4L+2}$] (d4Lp6) at (-4,0) {};
\node[gaugeb, label=above:$C_{2k+5L+1}$] (c5Lp6) at (-5,0) {};
\node[gauger, label=below:$D_{2k+6L+1}$] (d6Lp7) at (-6,0) {};
\node[gaugeb, label=above:$C_{k+4L}$] (c4Lp4) at (-7,0) {};
\node[gauger, label=below:$D_{2L}$] (d2Lp2) at (-8,0) {};
\node[gaugeb, label=above:$C_{k+3L+3}$] (c3Lp3) at (-6,1) {};

\draw[-] (d4)--(cLp4)--(d2Lp5)--(c3Lp5)--(d4Lp6)--(c5Lp6)--(d6Lp7)--(c4Lp4)--(d2Lp2) (d6Lp7)--(c3Lp3);

\node[gaugeb, label=left:$C_{n_1}$] (cn1) at (-1,1){};

\node (cdots) at (0,1) {$\cdots$};
\node[gaugeb, label=right:$C_{n_j}$] (cnj) at (1,1){};

\draw[-] (cn1)--(d4)--(cnj);

\draw (cn1) to [out=135, in=45,looseness=8] (cn1);
\draw (cnj) to [out=135, in=45,looseness=8] (cnj);

\end{tikzpicture}}
\end{gathered}\end{equation}Let $T=L+R$ denote the total number of $\mathrm{NS}5$ branes that move onto $\mathrm{O8}^-$ planes, then the total number of magnetic quivers is \begin{equation}
    \textrm{Number of magnetic quivers}=\sum_{T=0}^{N+2}\left\lceil\frac{T+1}{2}\right\rceil p(N+2-T)
\end{equation}where $p(N+2-T)$ denotes the total number of partitions of $N+2-T$.

It appears that the magnetic quiver in \eqref{quiver:magneticLSTNKGeneric} is somewhat overspecified given that it takes four parameters and a partition to define, whereas the 6d theory only takes two! In practise, not all of the choices of parameters will give computable magnetic quivers despite the non-negative balance of the gauge nodes.

Although each phase of the 6d theory and the corresponding Higgs branch -- described by the most generic magnetic quiver -- are interesting in their own right, it is helpful to restrict to a special case here.

The special case under consideration is $T=N+2$, which places all $\mathrm{NS}5$ branes on the $\mathrm{O8}^-$ planes. This leads to the magnetic quiver \qquiver[quiver:magneticLSTNK]{LST}{N, k,L} of the $\mathrm{LST}_{N,k}$ theory:

\begin{equation}\label{quiver:magneticLSTNK}
\begin{gathered}
\resizebox{\linewidth}{!}{\begin{tikzpicture}
\node[gauger, label=below:$D_{2k+4}$] (d4) at (0,0) {};
\node[gaugeb, label=above:$C_{2k+R+3}$] (cRp4) at (1,0) {};
\node[gauger, label=below:$D_{2k+2R+3}$] (d2Rp5) at (2,0) {};
\node[gaugeb, label=above:$C_{2k+3R+2}$] (c3Rp5) at (3,0) {};
\node[gauger, label=below:$D_{2k+4R+2}$] (d4Rp6) at (4,0) {};
\node[gaugeb, label=above:$C_{2k+5R+1}$] (c5Rp6) at (5,0) {};
\node[gauger, label=below:$D_{2k+6R+1}$] (d6Rp7) at (6,0) {};
\node[gaugeb, label=above:$C_{k+4R}$] (c4Rp4) at (7,0) {};
\node[gauger, label=below:$D_{2R}$] (d2Rp2) at (8,0) {};
\node[gaugeb, label=above:$C_{k+3R}$] (c3Rp3) at (6,1) {};

\draw[-] (d4)--(cRp4)--(d2Rp5)--(c3Rp5)--(d4Rp6)--(c5Rp6)--(d6Rp7)--(c4Rp4)--(d2Rp2) (d6Rp7)--(c3Rp3);

\node[gaugeb, label=above:$C_{2k+L+3}$] (cLp4) at (-1,0) {};
\node[gauger, label=below:$D_{2k+2L+3}$] (d2Lp5) at (-2,0) {};
\node[gaugeb, label=above:$C_{2k+3L+2}$] (c3Lp5) at (-3,0) {};
\node[gauger, label=below:$D_{2k+4L+2}$] (d4Lp6) at (-4,0) {};
\node[gaugeb, label=above:$C_{2k+5L+1}$] (c5Lp6) at (-5,0) {};
\node[gauger, label=below:$D_{2k+6L+1}$] (d6Lp7) at (-6,0) {};
\node[gaugeb, label=above:$C_{k+4L}$] (c4Lp4) at (-7,0) {};
\node[gauger, label=below:$D_{2L}$] (d2Lp2) at (-8,0) {};
\node[gaugeb, label=above:$C_{k+3L}$] (c3Lp3) at (-6,1) {};

\draw[-] (d4)--(cLp4)--(d2Lp5)--(c3Lp5)--(d4Lp6)--(c5Lp6)--(d6Lp7)--(c4Lp4)--(d2Lp2) (d6Lp7)--(c3Lp3);
\end{tikzpicture}}
\end{gathered}
% \begin{gathered}
%     \includegraphics[page=3,width=0.85\linewidth]{HSDoubling_Figures.pdf}
% \end{gathered}   
\end{equation}
where $R=N+2-L$ and so the magnetic quiver \qquiver[quiver:magneticLSTNK]{LST}{N, k,L} is a three parameter family depending on the triple $(N,k,L)$.

In the cases where Coulomb branch Hilbert series computations are possible, four $\orm(8)$ symmetries appear in the adjoint action, signalling the presence of a double $\orm(8)\times \orm(8)$ symmetry undergoing mitosis. This is because the magnetic quiver has a set of balanced gauge nodes going from $C_{2k+L+3}$ to the fork of $C_{k+3L}$ and $C_{k+4L}$ giving $\orm(8)\times \orm(8)$ and another factor swapping $L$ and $R$. The result is in agreement with the $\mathbb{Z}_2\times \mathbb{Z}_2$ automorphism of the curve configuration in \eqref{eqn:lstso8so8curveNK}. In the $N=0$ case the symmetry of the curve configuration corresponding to $\mathrm{LST}_{0,k}$ naïvely enhances to $S_4$,\footnote{A more accurate geometrical analysis should see an $S_2\times S_2$ symmetry action. This can be understood by thinking of the M-theory realisation of this LST having two distinguished M9-walls \cite{DelZotto:2022ohj}.} realising in the 3d magnetic theory a doubly 2-mitotic $\orm(8)$ theory.

\subsection{Two Heterotic $\mathrm{Spin}(32)$ Instantons}
Interestingly, the \qquiver[quiver:magneticLSTNK]{LST}{0, 0,1} theory is not the lowest-rank example within this family. Setting $N=0$, $k=-1$, and $L=1$ also yields a computable magnetic quiver:
\begin{equation}
\begin{gathered}
   \resizebox{0.85\linewidth}{!}{\begin{tikzpicture}
         \node[gaugeb, label=right:$C_2$] (c6t) {};
        \node[gauger, label=below:$D_{5}$] (d13) [below=of c6t]{};
        \node[gaugeb, label=below:$C_{4}$] (c11) [left=of d13]{};
        \node[gauger, label=below:$D_{4}$] (d10) [left=of c11]{};
        \node[gaugeb, label=below:$C_3$] (c8) [left=of d10]{};
        \node[gauger, label=below:$D_3$] (d7) [left=of c8]{};
        \node[gaugeb, label=below:$C_2$] (c5) [left=of d7]{};
        \node[gauger, label=below:$D_2$] (d4) [left=of c5]{};
        
        \node[gaugeb, label=below:$C_3$] (c8r) [right=of d13]{};
        \node[gauger, label=below:$D_2$] (d4r) [right=of c8r]{};

        \draw[-] (d4)--(c5)--(d7)--(c8)--(d10)--(c11)--(d13)--(c8r)--(d4r) (d13)--(c6t);

        \node[] (ghostb) [below=of d4]{};
        \node[] (ghostt) [above=of d4]{};

         \draw[dashed,red] (ghostb)--(d4.south) (d4.north)--(ghostt);
    \end{tikzpicture}}\label{quiv:mystery} 
\end{gathered}
\end{equation}
where the vertical dashed line is the usual symmetry axis along which the quiver is reflected. The finite coupling magnetic quiver given in \eqref{quiver:magneticLSTNKFinite} is not valid for this choice of triple due to negative rank gauge nodes. Taking the 6d electric quiver \Quiver{eq:LSTNK} and setting $N=0$ and $k=-1$ gives the following electric quiver \Quiver{quiv:Sp-1} in \eqref{quiv:Sp-1}. The $\sprm(-1)$ has had an unclear interpretation in 6d field theory, however the proposed interpretation of an $\sprm(-1)$ gauge group is as bi-spinor matter of $\orm(a)$ and $\orm(12-a)$ for $a=1,2,\cdots,12$. With this interpretation, the ``Lagrangian'' electric quiver \Quiver{quiv:NoSp-1} is given in \eqref{quiv:NoSp-1}.

\begin{equation}
\begin{gathered}
    \begin{tikzpicture}
    \node[gauger,label=below:{\footnotesize $\orm(4)$}] at (0,0) (c) {};
    \node[gaugeb,label=above:{\footnotesize $\sprm(-1)$}] [above right=0.6cm and 0.6cm of c] (ar) {};
    \node[gaugeb,label=below:{\footnotesize $\sprm(-1)$}] [below right=0.6cm and 0.6cm of c] (br) {};
    \node[gaugeb,label=above:{\footnotesize $\sprm(-1)$}] [above left=0.6cm and 0.6cm of c] (al) {};
    \node[gaugeb,label=below:{\footnotesize $\sprm(-1)$}] [below left=0.6cm and 0.6cm of c] (bl) {};
    \node[flavourr,label=right:{\footnotesize $\orm(8)$}] [right=0.8cm of ar] (far) {};
    \node[flavourr,label=right:{\footnotesize $\orm(8)$}] [right=0.8cm of br] (fbr) {};
    \node[flavourr,label=left:{\footnotesize $\orm(8)$}] [left=0.8cm of al] (fal) {};
    \node[flavourr,label=left:{\footnotesize $\orm(8)$}] [left=0.8cm of bl] (fbl) {};
    \draw[-,red] (fbl)--(bl) (fal)--(al) (fbr)--(br) (far)--(ar) (bl)--(c) (al)--(c) (br)--(c) (ar)--(c);
    \end{tikzpicture}
\label{quiv:Sp-1}
\end{gathered}
\end{equation}

\begin{equation}
\begin{gathered}
    \begin{tikzpicture}
    \node[gauger,label=below:{\footnotesize $\orm(4)$}] at (0,0) (c) {};
    \node[flavourr,label=above:{\footnotesize $\orm(8)$}] [above right=0.6cm and 0.6cm of c] (ar) {};
    \node[flavourr,label=below:{\footnotesize $\orm(8)$}] [below right=0.6cm and 0.6cm of c] (br) {};
    \node[flavourr,label=above:{\footnotesize $\orm(8)$}] [above left=0.6cm and 0.6cm of c] (al) {};
    \node[flavourr,label=below:{\footnotesize $\orm(8)$}] [below left=0.6cm and 0.6cm of c] (bl) {};
    
    \draw[-,red] (bl)--(c) (al)--(c) (br)--(c) (ar)--(c);
    \end{tikzpicture}
\label{quiv:NoSp-1}
\end{gathered}
\end{equation}

Considering the magnetic quiver \qquiver[quiver:magneticLSTNK]{LST}{0, -1, 1} in \eqref{quiv:mystery} for now, the HS is computed as \begin{align}
    \hsCLST{quiver:magneticLSTNK}{LST}{0,-1,1}&=1 + 992 t^2 + 419552 t^4 \nonumber\\&+ 102121006 t^6 + 16428483760 t^8 \nonumber\\&+ 
 1897591910432 t^{10} \nonumber\\&+ 166408046495505 t^{12}+O(t^{13})\\&=\big(1 + 496 t^2 + 86768 t^4 + 8023575 t^6 \nonumber\\&+ 470205768 t^8 + 
 19384338688 t^{10} \nonumber\\&+ 601699290368 t^{12}+O(t^{13})\big)^2\\
 \mathrm{PL}\left[\hsCLST{quiver:magneticLSTNK}{LST}{0,-1,1}\right]&=992 t^2 - 72976 t^4 + 11322254 t^6 \nonumber\\&- 2116928000 t^8 + 
 428647933824 t^{10} \nonumber\\&- 90783093942904 t^{12}+O(t^{13})
\end{align}
The moduli space is identified as the product moduli space \begin{equation}
    \mathcal C\left(\text{\qquiver[quiver:magneticLSTNK]{LST}{0, -1,0}}\right)=\left(\overline{min. \sorm(32)}\right)^2
\end{equation}
or equivalently as the product of two (reduced) moduli space of $\mathrm{Spin}(32)$ instantons on $\mathbb C^2$ with global symmetry $\mathrm{Spin}(32)/(\mathbb Z_2\times\mathbb Z_2)\times\mathrm{Spin}(32)/(\mathbb Z_2\times\mathbb Z_2)$. The full moduli space of one $G$ instanton on $\mathbb C^2$ also includes a free non-singular factor of $\mathbb H$.

Taking the interpretation of $\sprm(-1)$ gauge group as having a D6 of charge $-1$ on $\mathrm{O6}^+$, the following Type I$'$ brane system realises the electric quiver \Quiver{quiv:Sp-1} in the electric phase and the magnetic quiver \qquiver[quiver:magneticLSTNK]{LST}{0, -1,0} in the magnetic phase.
\begin{equation}
    \begin{gathered}
        \resizebox{0.85\linewidth}{!}{\begin{tikzpicture}

        \ns{2,0};
        \ns{4,0};
        \draw[-] (1,-1)--(1,1)node[pos=0,below]{$8$} (5,-1)--(5,1)node[pos=0,below]{$8$};
        \draw[dashed] (6,-1)--(6,1) node[pos=1,above]{$\mathrm{O8}^-$};
        \draw[dashed] (0,-1)--(0,1) node[pos=1,above]{$\mathrm{O8}^-$};
        \node[circle, draw, fill=white, label=right:$\mathrm{ON}^{0}$] at (6,0) 
        {};
        \node[circle, draw, fill=white, label=left:$\mathrm{ON}^{0}$] at (0,0) 
        {};
        
        \draw[-] (2,0.17)--(4,0.17);
        \draw[-] (2,-0.17)--(4,-0.17);
        \draw[-,red] (0,0.17)--(2,0.17) (4,0.17)--(6,0.17);
        \draw[-,red] (0,-0.17)--(2,-0.17) (4,-0.17)--(6,-0.17);

        \node[label=above:$2$] at (3,0){};
        \node[label=above:$-1$] at (1.5,0){};
        \node[label=above:$-1$] at (4.5,0){};
        \node[label=below:$-1$] at (1.5,0){};
        \node[label=below:$-1$] at (4.5,0){};
        
        \draw[dotted] (0.2,0)--(1.8,0) (4.2,0)--(5.8,0);
    \end{tikzpicture}}\label{eq:SO32brane}
    \end{gathered}
\end{equation}

\section{Higgs branch Hasse Diagrams and Mitosis}\label{sec:doubling}

The previous sections contain a number of quivers exhibiting a phenomenon termed `mitosis'; the aim of this section is to motivate the use of this label as an addition to the decay and fission algorithm \cite{Lawrie:2024wan}. Symmetry mitosis is broadly defined as the appearance of \emph{two} identical global symmetry factors $F$ in the Coulomb branch of a 3d $\mathcal{N}=4$ quiver, arising from a common set of balanced nodes that bifurcate into two distinct chains (see Appendix~\ref{sec:Appendix}). Each of these symmetry factors sources an equivalent but distinguishable direction in the Hasse diagram, creating a diamond structure or otherwise `multiplying' the Hasse diagram from the bottom-up. \footnote{\label{fn:instanton}The doubling effect is significantly complicated by instanton moduli space structures in the Hasse diagram. Such objects will not be considered in detail here.} Higher bifurcations are not currently understood (consider the discussion of \eqref{quiver:multi-leg}) and remain an open question.

\subsection*{Hasse Diagram Diamonds and Symmetry Mitosis}
The phenomenon of Hasse diagram doubling is one of the simplest consequences of symmetry mitosis in a magnetic quiver. Consider again the example of the magnetic quiver \eqref{quiver:magneticOINK} at $N=1$ and $k=0$. The $S_3$ permutation of the three $\orm(8)$ flavour symmetries in the electric theory gives (starting at the bottom leaf) three identical Higgsing transitions in the Hasse diagram. The decay and fission algorithm \cite{Lawrie:2024wan} however only sees two, which is incorrect. Na\"ive use of quiver subtraction also does not make manifest the three $\orm(8)$. The third arises due to the fact that two of the $\orm(8)$ symmetries in the magnetic quiver come from the bifurcation at the $\mathfrak{d}_{13}$ gauge node, contributing two balanced chains instead of one. The Hasse diagram for the magnetic quiver hence has two identical $\overline{min.D_4}$ transitions emerging from the lowest node in its Hasse diagram, and extra one coming from the remaining $\orm(8)$ higgsing.

\begin{figure}[h!]
\includegraphics[width=0.6\linewidth,page=7]{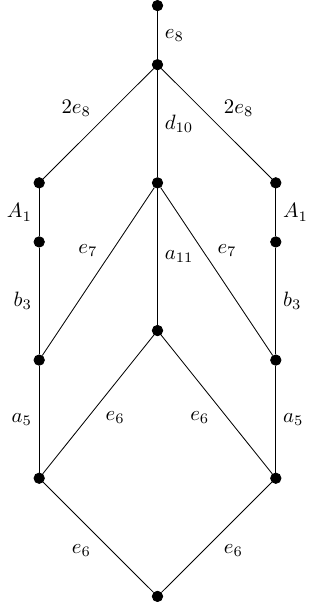}

    \caption{Higgs branch Hasse diagram for $(E_6,E_6)$ minimal conformal matter.}
    \label{fig:E6E6HasseClean}
\end{figure}

The same phenomenon arises in the magnetic quiver \eqref{quiv:E6E6Magnetic} with mitotic $E_6\times E_6$ global symmetry. Here the decay and fission algorithm would only see one $E_6$ Higgsing emerging from the bottom node, as would naïve quiver subtraction -- the mitosis instead causes this to be doubled, agreeing with partial predictions from 6d \cite{Bao:2024eoq,Bao:2025pxe}. The full Hasse diagram with symmetry mitosis accounted for is given in \Figref{fig:E6E6HasseClean}.

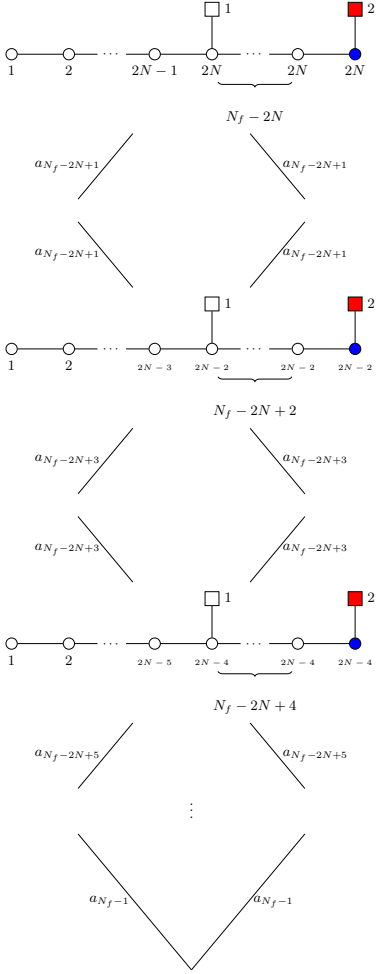
\begin{figure}
    \centering
    \resizebox{!}{13cm}{
    \begin{tikzpicture}
        \node (a) at (0,0) {$\begin{tikzpicture}
            \node[gaugeb,label=below:{$2N$}] at (0,0) (usp) {};
            \node[flavourr,label=right:{$2$}] [above=0.7cm of usp] (f1) {};
            \node[gauge,label=below:{$2N$}] [left=1cm of usp] (u2n) {};
            \node [left=0.5cm of u2n] (dots) {\footnotesize $\cdots$};
            \node[gauge,label=below:{$2N$}] [left=0.5cm of dots] (u2nf) {};
            \node[flavour,label=right:{$1$}] [above=0.7cm of u2nf] (f2) {};
            \node[gauge,label=below:{$2N-1$}] [left=1cm of u2nf] (u2n-1) {};
            \node [left=0.5cm of u2n-1] (dots2) {\footnotesize $\cdots$};
            \node[gauge,label=below:{$2$}] [left=0.5cm of dots2] (u2) {};
            \node[gauge,label=below:{$1$}] [left=1cm of u2] (u1) {};
            \draw[decorate,decoration={brace, amplitude=3pt, raise=4.5ex}] (u2n.west)--(u2nf.east) node[midway,label={[yshift=-50pt]$N_f-2N$}] () {};
            \draw (f1)--(usp) (usp)--(u2n) (u2n)--(dots) (dots)--(u2nf) (u2nf)--(f2) (u2nf)--(u2n-1) (u2n-1)--(dots2) (dots2)--(u2) (u2)--(u1);
        \end{tikzpicture}$};
        \node (b) at (0,-6.5) {$\begin{tikzpicture}
            \node[gaugeb,label=below:{\tiny$2N-2$}] at (0,0) (usp) {};
            \node[flavourr,label=right:{$2$}] [above=0.7cm of usp] (f1) {};
            \node[gauge,label=below:{\tiny$2N-2$}] [left=1cm of usp] (u2n) {};
            \node [left=0.5cm of u2n] (dots) {\footnotesize $\cdots$};
            \node[gauge,label=below:{\tiny$2N-2$}] [left=0.5cm of dots] (u2nf) {};
            \node[flavour,label=right:{$1$}] [above=0.7cm of u2nf] (f2) {};
            \node[gauge,label={below:\tiny$2N-3$}] [left=1cm of u2nf] (u2n-1) {};
            \node [left=0.5cm of u2n-1] (dots2) {\footnotesize $\cdots$};
            \node[gauge,label=below:{$2$}] [left=0.5cm of dots2] (u2) {};
            \node[gauge,label=below:{$1$}] [left=1cm of u2] (u1) {};
            \draw[decorate,decoration={brace, amplitude=3pt, raise=4.5ex}] (u2n.west)--(u2nf.east) node[midway,label={[yshift=-50pt]$N_f-2N+2$}] () {};
            \draw (f1)--(usp) (usp)--(u2n) (u2n)--(dots) (dots)--(u2nf) (u2nf)--(f2) (u2nf)--(u2n-1) (u2n-1)--(dots2) (dots2)--(u2) (u2)--(u1);
        \end{tikzpicture}$};

        \node (c) at (0,-13) {$\begin{tikzpicture}
            \node[gaugeb,label=below:{\tiny$2N-4$}] at (0,0) (usp) {};
            \node[flavourr,label=right:{$2$}] [above=0.7cm of usp] (f1) {};
            \node[gauge,label=below:{\tiny$2N-4$}] [left=1cm of usp] (u2n) {};
            \node [left=0.5cm of u2n] (dots) {\footnotesize $\cdots$};
            \node[gauge,label=below:{\tiny$2N-4$}] [left=0.5cm of dots] (u2nf) {};
            \node[flavour,label=right:{$1$}] [above=0.7cm of u2nf] (f2) {};
            \node[gauge,label={below:\tiny$2N-5$}] [left=1cm of u2nf] (u2n-1) {};
            \node [left=0.5cm of u2n-1] (dots2) {\footnotesize $\cdots$};
            \node[gauge,label=below:{$2$}] [left=0.5cm of dots2] (u2) {};
            \node[gauge,label=below:{$1$}] [left=1cm of u2] (u1) {};
            \draw[decorate,decoration={brace, amplitude=3pt, raise=4.5ex}] (u2n.west)--(u2nf.east) node[midway,label={[yshift=-50pt]$N_f-2N+4$}] () {};
            \draw (f1)--(usp) (usp)--(u2n) (u2n)--(dots) (dots)--(u2nf) (u2nf)--(f2) (u2nf)--(u2n-1) (u2n-1)--(dots2) (dots2)--(u2) (u2)--(u1);
        \end{tikzpicture}$};
        
        \draw[-] (a) -- (2.5,-3) node[pos=0.5, right]{$a_{N_f-2N+1}$};
        \draw[-] (2.5,-3.5) -- (b) node[pos=0.5, right]{$a_{N_f-2N+1}$};
        \draw[-] (a) -- (-2.5,-3) node[pos=0.5, left]{$a_{N_f-2N+1}$};
        \draw[-] (-2.5,-3.5) -- (b) node[pos=0.5, left]{$a_{N_f-2N+1}$};

        \draw[-] (b) -- (2.5,-9.5) node[pos=0.5, right]{$a_{N_f-2N+3}$};
        \draw[-] (2.5,-10) -- (c) node[pos=0.5, right]{$a_{N_f-2N+3}$};
        \draw[-] (b) -- (-2.5,-9.5) node[pos=0.5, left]{$a_{N_f-2N+3}$};
        \draw[-] (-2.5,-10) -- (c) node[pos=0.5, left]{$a_{N_f-2N+3}$};
        \draw[-] (c) -- (2.5,-16) node[pos=0.5, right]{$a_{N_f-2N+5}$};
        \draw[-] (c) -- (-2.5,-16) node[pos=0.5, left]{$a_{N_f-2N+5}$};
        \node (5) at (0, -16.5) {$\rvdots$};
        \draw[-] (2.5,-17)  -- (0,-20) node[pos=0.5, right]{$a_{N_f-1}$};
        \draw[-] (-2.5,-17) -- (0,-20) node[pos=0.5, left]{$a_{N_f-1}$};
    \end{tikzpicture}
    }
    \caption{The Coulomb branch Hasse diagram for \eqref{eqn:magnetic_quiver2U} consists of diamonds of minimal nilpotent orbit closures of $A$-type Lie algebras. Note that only the bottom/top leaves of a diamond can be associated with the Coulomb branch of a magnetic quiver --- an artifact of the $\mathrm{ON}^{0}$ plane in the brane system.}
    \label{fig:A_nA_n_CBHD}
\end{figure}

Consider again the brane system in \eqref{eqn:electric-branes+quiver2U(N)} and its corresponding electric and magnetic theories in \eqref{eqn:magnetic_quiver2U}. As previously stated, the $\surm(k)\times \surm(k)$ global symmetry of the Coulomb branch of the unitary-orthosymplectic theory \eqref{eqn:magnetic_quiver2U} stems from the $N+1$ balanced set of nodes ranging from the $\urm(1)$ to the $C_{N}$ nodes. Notice that this single chain sources both symmetry factors, consistent with expectations from symmetry mitosis. The usual balance rules predict the $\surm(k)$ factor in the chain involving the unitary nodes only. Higgsing only a single factor of the global symmetry is unclear, but since the two symmetries are intertwined a combined Higgsing of both the symmetry factors is possible. In the 3d mirror theory Higgsing a single $\surm(k)$ factor is still physical. A Hasse diagram for the quiver theory \eqref{eqn:magnetic_quiver2U} must display this alternative Higgsing, as depicted in Figure
\ref{fig:A_nA_n_CBHD}. Hence while quiver subtraction reproduces the correct Hasse diagram in \Figref{fig:A_nA_n_CBHD}, subtraction quivers are not known for all slices.

\begin{figure*}
    \centering

\includegraphics[width=0.9\textwidth,page=4]{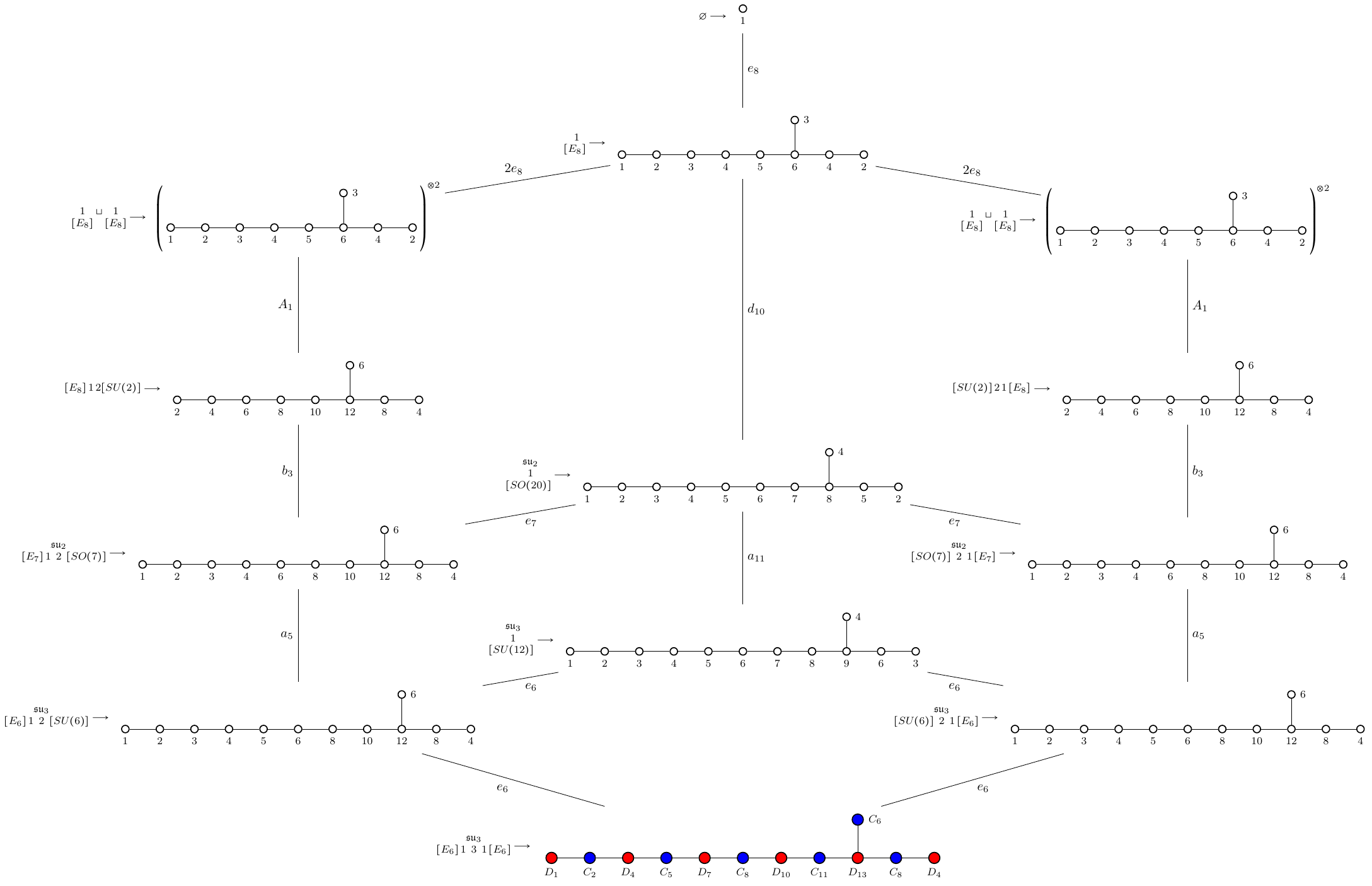}

    \caption{Hasse diagram for the Coulomb branch of the magnetic quiver \qquiver[quiv:E6E6Magnetic]{OI}{E_6} with the respective 6d electric theories.}
    \label{fig:E6E6HasseDF}
\end{figure*}
%consistently with the presence of a $3$-mitosis of the $\orm(8)$ symmetry in \qquiver[quiver:magneticOINK]{OI}{1,0}.

In general, mitotic symmetries in unframed orthosymplectic quivers lead to a multiplicity in the Hasse diagram transitions that arise from that symmetry. Given the valence-three bifurcations in many of the quivers studied in this note, their associated transitions in the Hasse diagram will be doubled. This operation must be applied \emph{after} the decay and fission algorithm in order to capture all the phases of the theory. In the decay and fission algorithms, any decay of gauge nodes away from the bifurcation are doubled whereas decays of the gauge nodes in the bifurcation are not doubled. This is due to the $\mathrm{ON}^0$ plane in the corresponding brane system, which gives doubled transitions for branes away from the orientifold and one transition for branes which make direct contact with it.

\section{Conclusions and Outlook}

This note defines the phenomenon of symmetry mitosis in unframed orthosymplectic magnetic quivers and explores examples from 3d product theories in Section~\ref{sec:Mitotic_Unitary}, and 6d SQFTs in Section~\ref{sec:Mitotic6dSQFT} and Section~\ref{sec:Mitotic6dLST}.\\
The observations made in this paper identify distinct Higgsing directions in the Coulomb branch Hasse diagram associated to a magnetic quiver. These directions are determined by discrete symmetry factors acting on the local geometry of the higher-dimensional electric theory, as shown in Section~\ref{sec:Mitotic6dSQFT} and Section~\ref{sec:Mitotic6dLST}. As such, this leads to the prospect of encoding discrete symmetries inside the magnetic quiver procedure, connecting thus the naïve algebra-group correspondence to the specification of a more precise global form assignment for the group. Various subtleties emerge here, including those concerning instanton moduli spaces remarked on in Footnote~\ref{fn:instanton}. In the moduli space of two $E_8$ instantons there exists a one quaternionic-dimensional Higgsing transition which makes them coincident, leading to a product theory where each side can be completely Higgsed. Of course, this is not a strict product, as the theory is unable to distinguish which $E_8$-shaped copy is being acted upon and the associated slice in the Hasse diagram is a union of two minimal $E_8$ nilpotent orbits.
% \begin{figure}[h!]
%     \centering   \includegraphics[page=5,width=0.6\linewidth]{HSDoubling_Figures.pdf}
%     \caption{\textcolor{red}{CHANGE}Coulomb branch Hasse Diagram the 3d $\mathcal{N}=4$ quiver gauge theory whose Colomb branch is the moduli space of two $E_8$ instantons.}
%     \label{fig:2E8-instanton}
% \end{figure}
It would be interesting to understand the precise relation between this and the mitotic orthosymplectic quiver whose Coulomb branch is the product of minimal nilpotent orbit closures.
\begin{figure}[h!]
    \centering   \includegraphics[page=6,width=0.6\linewidth]{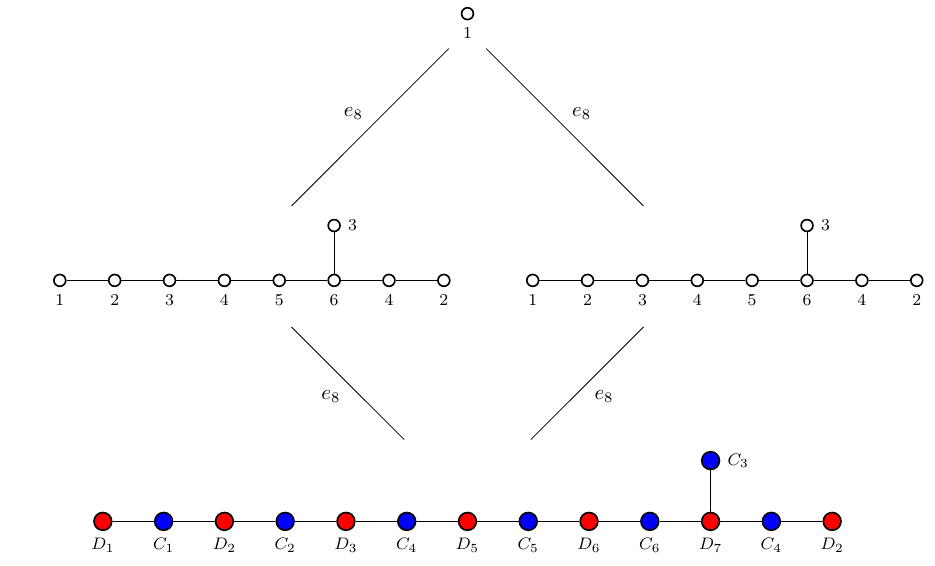}
    \caption{Coulomb branch Hasse Diagram the 3d $\mathcal{N}=4$ quiver gauge theory exhibiting $E_8$ 2-mitosis.}
    \label{fig:2-mitosisE8}
\end{figure}\\

\textbf{Outlook.} It is possible to imagine a star-shaped quiver with a balanced central node and $n>2$ emergent legs. It turns out that only one family of theories, given in  \eqref{quiver:multi-leg}, ensure that all nodes connected to the central node are balanced and the realised theory is good \cite{Gaiotto:2008ak}.

Perturbative Hilbert Series computations fail to produce concrete results and hence the existence of a 3-mitosis cannot be discounted. Confirming this phenomenon involves exploiting the fact that such 3-mitoses may occur in magnetic quivers of certain 4d $\mathcal{N}=2$ Class $\mathcal{S}$ theories \cite{Chacaltana:2011ze, Gaiotto:2009hg}. A Hall-Littlewood index computation confirms that the theory so realised is bad \cite{Gaiotto:2008ak}.

\section*{Acknowledgements}
We would like to thank Michele Del Zotto, Jacques Distler, Craig Lawrie, and Marcus Sperling for interesting discussions and useful comments. The work of SB, AH, and GK is partially supported by STFC Consolidated Grants ST/T000791/1 and ST/X000575/1. The work of SB is supported by the STFC DTP research studentship grant ST/Y509231/1. The work of GK is supported by STFC DTP research studentship grant ST/X508433/1.
LM~thanks Imperial College London for hospitality during the initial phase of this work. LM~acknowledges support from DESY (Hamburg, Germany), a member of the Helmholtz Association HGF; LM~also acknowledges the Deutsche Forschungsgemeinschaft under Germany's Excellence Strategy - EXC 2121 ``Quantum Universe'' - 390833306 and the Collaborative Research Center - SFB 1624 ``Higher Structures, Moduli Spaces, and Integrability'' - 506632645.

\appendix
\section*{Balance Rules in Quiver Gauge Theories}\label{sec:Appendix}
The notion of balance \cite{Gaiotto:2008ak} in a 3d$\;\mathcal N=4$ theory is defined in terms of the R-charges of monopole operators. In a quiver gauge theory without adjoint hypermultiplets the balance $b$ is computed using the rank of the gauge factors. The formulae for $\urm(n), \;\orm(n),$ and $\sprm(n)$ gauge nodes are given below.

For unitary quiver gauge theories the balance of an $\urm(n)$ gauge node which sees $k$ flavours is \begin{equation}\label{eqn:balanceU}
    b=k-2n.
\end{equation}

For an $\orm(n)$ gauge node that sees $k$ $\sprm$ fundamental hypermultiplets the balance is \begin{equation}\label{eqn:balanceSO}
    b=k-n+1.
\end{equation}

For an $\sprm(n)$ gauge node which sees $k$ $\orm$ fundamental hypermultiplets the balance is \begin{equation}\label{eqn:balanceSp}
    b=k-2n-1.
\end{equation}

The balance of a gauge nodes affects the convergence of the Hilbert series when computed with the monopole formula. There are three cases to consider: \begin{itemize}
    \item All gauge nodes have $b>0$ --- The monopole formula usually converges.
    \item All gauge nodes have $b>0$ except one with $b=-1$ --- The monopole formula converges for unitary quivers and not for orthosymplectic.
    \item Any gauge node has balance $b\leq-2$ or multiple gauge nodes have balance $b=-1$ --- The monopole formula never converges.
\end{itemize}

The Coulomb branch global symmetry at a generic point in the moduli space is $\urm(1)^r$ where $r$ is the rank of the gauge group. Depending on the details of the theory, this global symmetry may enhance to some non-Abelian group. In a quiver gauge theory this enhancement typically occurs when there are multiple gauge nodes which are balanced, $b=0$. In unitary quivers any subset of balanced gauge nodes forming a $G$ Dynkin diagram typically corresponds to a factor of $G$ in the Coulomb branch global symmetry. For orthosymplectic quivers this is more subtle. A chain of $p$ balanced gauge nodes typically gives the following contributions to the global symmetry:\begin{itemize}
    \item $\orm(p+1)$ if there are no $D_1$ gauge nodes in the chain.
    \item $\orm(p+2)$ if there is a $D_1$ gauge nodes at one end of the chain.
    \item $\orm(p+3)$ if there are $D_1$ gauge nodes at both ends of the chain.
\end{itemize}

Further enhancement of the global symmetry may occur from the half-integer magnetic lattice of the gauge groups \cite{Bourget2020MagneticQuivers}. If there are $l$ $D_n$ gauge nodes in the quiver (balanced or otherwise) then there are an additional $2^l$ generators, if they are at R-charge $1$ then these generators also contribute to the global symmetry.
\section*{Higher Mitoses}
Owing to the balance conditions on gauge nodes, it's straightforward to systematise higher mitoses in unframed orthosymplectic theories. Consider the junction in \eqref{quiver:higher-mitoses} below -- assuming that the $D_k$ and $C_{n^{i}_{1}}$ nodes are all balanced, and that $k$, $n_{j}^{i}\in\mathbb{Z}_{\geq0}$, the balance conditions given in \eqref{alignlevel1} and \eqref{alignlevel2} lead to \eqref{alignlevel3}.
\begin{equation}\label{quiver:higher-mitoses}
\begin{gathered}
 \begin{tikzpicture} 
        \node[gauger, label=below:$D_{k}$] (1) at (0,0){};
        \node[gaugeb, label=below:$C_{n_{1}^{1}}$] (2a) at (1,0){};
        \node[gauger, label=below:$D_{n_{2}^{1}}$] (3a) at (2,0){};
        \node[gaugeb, label=left:$C_{n_{1}^{\alpha}}$] (2b) at (0,1){};
        \node[gauger, label=left:$D_{n_{2}^{\alpha}}$] (3b) at (0,2){};
        \draw[-] (1)--(2a)--(3a)--(2.5,0) (1)--(2b)--(3b)--(0,2.5);
        \node[] (cdots) at (0.55,0.75) {$\ddots$};
    \end{tikzpicture}   
\end{gathered}
\end{equation}
\begin{align}
    2k-1 &= \sum_{i=1}^{\alpha}n^{i}_{1}\label{alignlevel1}\\
    2n^{i}_{1}+1&=k+n^{i}_{2}\label{alignlevel2}
\end{align}
This leads to the condition
\begin{equation}
    \alpha = 2\left(\frac{2k-1}{k-1}\right)-\frac{1}{k-1}\sum_{i=1}^{\alpha}n_{2}^{i}
\label{alignlevel3}
\end{equation}
This has three (non-trivial) families of solutions.
\begin{itemize}
    \item $k\in\mathbb{Z}_{\geq0}$ and $\sum_{i-1}^{\alpha}n_{2}^{i}=k+1$ gives $\alpha=3$. This is a standard 2-mitosis of the sort considered throughout this note.
    \item $k\in\mathbb{Z}$ and $\sum_{i-1}^{\alpha}n_{2}^{i}=2$ gives $\alpha=4$. This corresponds to quivers of the form \eqref{quiver:multi-leg}.
    \item $k=3$, $\sum_{i-1}^{\alpha}n_{2}^{i}=0$ and $\alpha=5$. This is the star-shaped quiver given by a central node of $D_{3}$ and five adjoining nodes of $C_{1}$. This was considered for instance in \cite{Lawrie:2024wan}.
\end{itemize}
\begin{equation}\label{quiver:multi-leg}
\begin{gathered}
 \begin{tikzpicture} 
        \node[gaugeb, label=below:$C_{n+1}$] (1C) {};
        \node[gauger, label=below:$D_{2n+1}$] (1D) [right=of 1C]{};
        \node[gauger, label=below:$D_{2}$] (2D) [left=of 1C]{};
        \node[gaugeb, label=right:$C_{n}$] (2C) [above=of 1D]{};
        \node[gaugeb, label=right:$C_{n}$] (3C) [above right=of 1D]{};
        \node[gaugeb, label=right:$C_{n}$] (4C) [right=of 1D]{};
        \node (Q) [draw, circle] at (-2.5,0) {$Q$};
        \draw[-] (Q)--(2D)--(1C)--(1D)--(2C)--(1D)--(3C)--(1D)--(4C);
    \end{tikzpicture}   
\end{gathered}
\end{equation}
Interestingly, higher mitoses are not possible for $D/C$ orthosymplectic quivers with a central node of $C$-type.

\bibliography{bibliography}
\end{document}